\documentclass[apl,preprint]{revtex4-1}
\usepackage{graphicx}
\usepackage{dcolumn} 
\usepackage{bm}
\usepackage{float}
\usepackage[normalem]{ulem}
\usepackage{color}

\begin{document}

\title{A study of electron and thermal transport in layered Titanium disulphide single crystals}
\author{Dhavala Suri$^1$, Vantari Siva$^2$, Shalikram Joshi$^3$, Kartik Senapati$^2$, P. K. Sahoo$^2$, Shikha Varma$^4$, R. S. Patel$^1$}
\affiliation{$^1$Department of Physics, Birla Institute of Technology \& Science  Pilani (BITS Pilani) - K K Birla Goa Campus,\\ Zuarinagar, Goa 403726, India}
\affiliation{$^2$School of Physical Sciences, National Institute of Science Education and Research, HBNI, Jatni, Odisha 752050, India}
\affiliation{$^3$Tata Steel, Jamshedpur, Jharkhand, India}
\affiliation{$^4$Institute of Physics (IOP), Bhubaneswar,
Odisha 751005, India}

\begin{abstract} 
We present a detailed study of thermal and electrical transport behavior of single crystal Titanium disulphide flakes, which belongs to the two dimensional, transition metal dichalcogenide class of materials. In-plane Seebeck effect measurements revealed a typical metal-like linear temperature dependence in the range of 85  - 285 K. Electrical transport measurements with in-plane current geometry exhibited a nearly $T^2$ dependence of  resistivity in the range of 10 - 300 K. However, transport measurements along the out-of-plane current geometry showed a transition in temperature dependence of resistivity from $T^2$  to $T^5$  beyond 200 K. Interestingly, Au ion-irradiated TiS$_2$ samples showed a similar $T^5$ dependence of resistivity beyond 200 K, even in the current-in-plane geometry.  Micro- Raman measurements were performed to study the phonon modes in both pristine and ion-irradiated TiS$_2$ crystals. 

\end{abstract}

\maketitle
\section{Introduction}

The discovery of graphene lead to an avalanche in research on two dimensional materials. Recently, the interest has focussed on  transitional metal dichalcogenides (TMDCs) which edge over graphene due to properties like non-saturating magnetoresistance, large Seebeck coefficient, high mobility and tunable band gap \cite{Snyder,radi,rscrev}. TMDCs are layered materials of the form MX$_2$, where M and X stand for transition metal and chalcogen, respectively. While the intra-layer bonding is strong and covalent in nature, the inter-layer bonding is through weak van der Waals forces. Hence these materials are easy to exfoliate and fabricate two dimensional devices. Confinement of electrons to two dimensions results in significant reduction of scattering and hence electronic and thermal conductivity differ remarkably from those in bulk. TMDCs are intriguing class of materials for both fundamental understanding of materials and technological applications. They possess high spin-orbit coupling and hence are useful in spin Hall effect based devices. They are found to exhibit superconducting phase, charge density wave phase etc. Signatures of Weyl semimetals have been observed in MoTe$_2$ and WTe$_2$  \cite{Jiang,Deng,Soluyanov,Lu2016}. Heterostructure devices of TMDCs with graphene, h-BN etc. enhance the spin dependent properties by proximity effect \cite{Avsar2014,Lee2013,Langouche2013}. In  device applications, field effect transistors of MoS$_2$ and MoSe$_2$ have displayed high on/off ratios of nearly 10$^8$ and 10$^6$ respectively \cite{radi,Larentis}. MoTe$_2$ in its semiconducting (2H) phase shows Seebeck coefficients of the order of mVK$^{-1}$.  WTe$_2$ and MoTe$_2$ have been reported to have a non-saturating extremely large magnetoresistance \cite{Ali2014,Keum2015}. Unlike graphene TMDCs have a finite tunable band gap in visible to near infrared range. In some of the TMDCs like MoS$_2$ drastic change in the electronic band structure with number of layers leads to a band gap cross-over from direct to indirect band gap. This property is utilized in photovoltaic and optoelectronic device applications \cite{Mak2016}.

In the family of TMDCs Titanium disulphide is also a potential candidate for thermopower device applications due to its low thermal conductivity and high Seebeck co-efficient. This material is known to be a semimetal in layered bulk form and a semiconductor in the monolayer form \cite{Fang1997}. It is extensively utilized in storage batteries as solid electrolyte \cite{battery}. It is also known to undergo semiconductor to metal phase transition under pressure \cite{Klipstein,Yu,ross}. However, detailed investigations of basic transport properties of this material are sparse in the literature. Klipstein et al., have discussed transport properties of TiS$_2$ and attribute the $T^2$ dependence of resistivity to inter-valley and intra-valley scattering \cite{Klip}. Imai et al., \cite{Imai} report ac resistivity measurements for in-plane and out of plane axes and observe $T^2$ variation of resistivity throughout the temperature and attribute this to inter-valley scattering. Theoretical investigations have explored electronic and optical properties and have explained charge density wave phase in TiS$_2$ \cite{Dolui}. Klipstein et al., provide a detailed analysis of  temperature dependence of resistivity with varying stoichiometry \cite{Klip}.  However, the  effects of structural anisotropy on the electrical properties and possibilities of tailoring those properties, need a closer inspection. This motivates us to conduct a detailed study on thermal and electrical transport properties of thin flakes of crystalline TiS$_2$. Particular emphasis was given to the anisotropic electrical properties via current-in-plane (CIP) and current-perpendicular to-plane (CPP) measurements. We discover an unconventional $T^5$  dependence of the out of plane electrical resistivity, at higher temperatures, which has not been observed prior to this report. Using ion irradiation induced defects, we were able to induce the $T^5$  dependence of resistivity in the CIP geometry as well.

\section{Experimental Details}

High purity single crystal TiS$_2$ oriented along (00l) planes were purchased from HQ Graphene. The hexagonal unit cell parameters as provided by HQ Graphene were, a = b =0.339 nm, c=~0.568 nm, $\alpha = \beta$ = 90$^\circ$, $\gamma$ =120$^\circ$. Samples used for our experiments were exfoliated from this bulk TiS$_2$.

\begin{figure}[h]
\centering
\includegraphics[width=8cm]{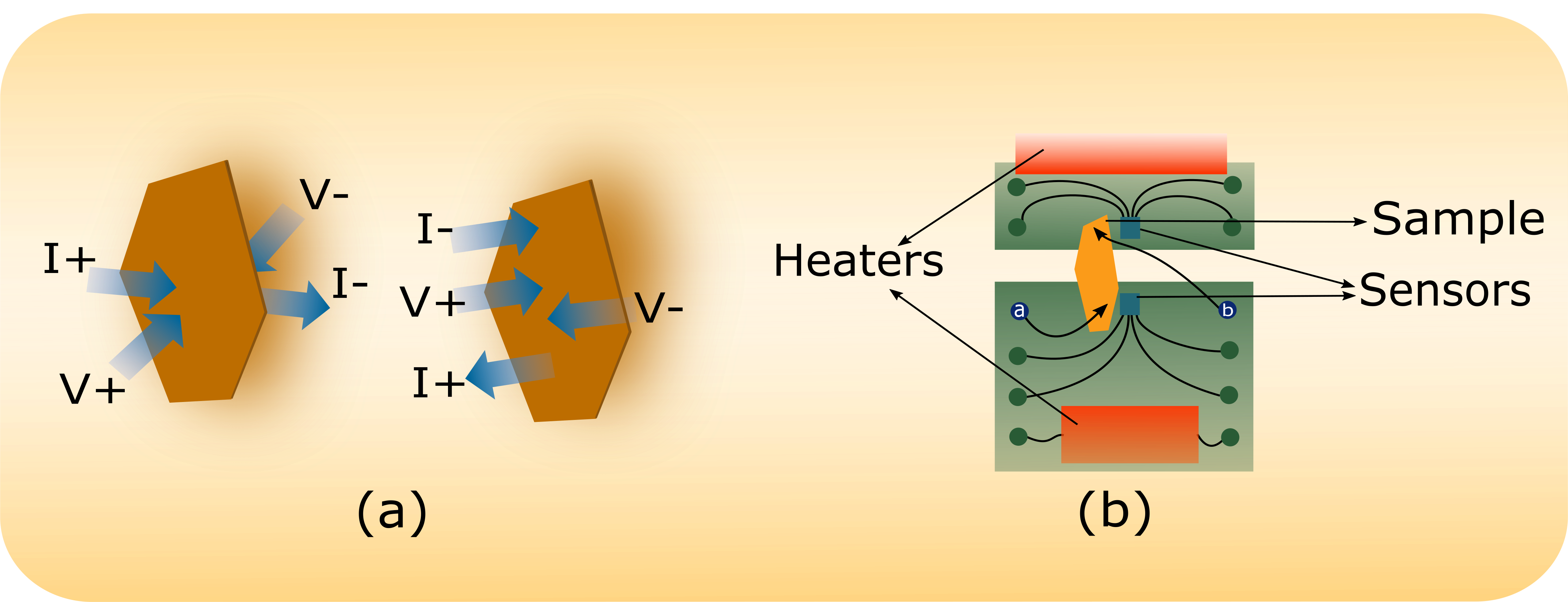}
\caption{Schematic for (a) CPP and CIP measurement modes (b) Seebeck measurement schematic.}
\end{figure}

For structural characterization XRD measurements were performed using Cu-K$\alpha$ radiation ($\lambda$ = 1.54 \AA) of Bruker D8 advanced diffractometer in $\theta - 2\theta$ configuration. Micro Raman scattering studies were carried out in back scattering geometry with Horiba Jobin Yvon spectrometer which is equipped with T64000 triple monochromator and liquid nitrogen cooled Charged Coupled Device (CCD) detector. 100$\times$ objective was used to acquire Raman spectra which provides 1 $\mu$m$^2$ laser spot on the sample surface. An Argon ion laser (514 nm) was used with very low power to avoid any damage to the sample due to laser heating.  Surface morphology was studied using Field Emission Scanning Electron Microscope (Carl Zeiss - FESEM). Nanoscope IIIA multimode (Veeco) was utilized for tapping mode Atomic Force Microscope (AFM) studies. Au ion irradiation was performed, with energy of 15 keV at normal incidence in high vacuum of the order of 10$^{-6}$ mbar. 

DC resistivity measurements were performed in CIP and CPP mode  using standard four probe technique. The typical sample dimensions were:  thickness 100 $\mu$m and area $5$ mm $\times$ $2$ mm. Contacts were made using silver paste on the same side of the sample for CIP measurements and on opposite sides for CPP mode (Fig. 1 (a)). Closed cycle He cryostat (Oxford Instruments) along with MercuryiTC temperature controller was used for resistivity measurements  as a function of temperature.   For Seebeck measurements the sample was placed between two thermal blocks (heaters) separated by an air gap of $1$ mm (Fig. 1 (b)). Copper wires were used as probes to measure thermal voltage. Contacts were made on the sample using silver paste. Temperature sensors were placed in close proximity to the probes to record the temperature. The sample was electrically insulated from the thermal block using mica sheet. The entire unit was mounted on the dip stick. The dip stick was dipped in a liquid nitrogen dewar for temperature dependent measurements. For both resistivity and Seebeck measurements Keithley nanovoltmeter and Keithely sourcemeter used as voltmeter and current source respectively. Lakeshore temperature controller was used for temperature control. The data acquisition for all the experiments  were LabVIEW automated.

\section{Results and Discussions}

\subsection{Structural analysis}
XRD spectrum indicates single crystal characteristics with sharp peaks along (001), (002), (003), (004) planes (Fig. 2).  Peaks at higher angles $47.5^\circ$ and $65.1^\circ$, show double lines corresponding to the K$_{\alpha1}$ and K$_{\alpha2}$ transitions. The wavelengths of K$_{\alpha1}$ and K$_{\alpha2}$ are 1.540 nm and 1.544 nm respectively which are extremely close. The peak at angle $47.5^\circ$ is shown explicitly in the inset, where we observe a doublet peak with lesser intensity at $47.65^\circ$ corresponding to the K$_{\alpha2}$ transition. The resolution of the doublet increases with angle  (2$\theta$). This implies a high quality of single crystalline nature of the sample. Inter-layer spacing is found to be 5.7 \AA.

\begin{figure}[H]
\centering
\includegraphics[width=8cm]{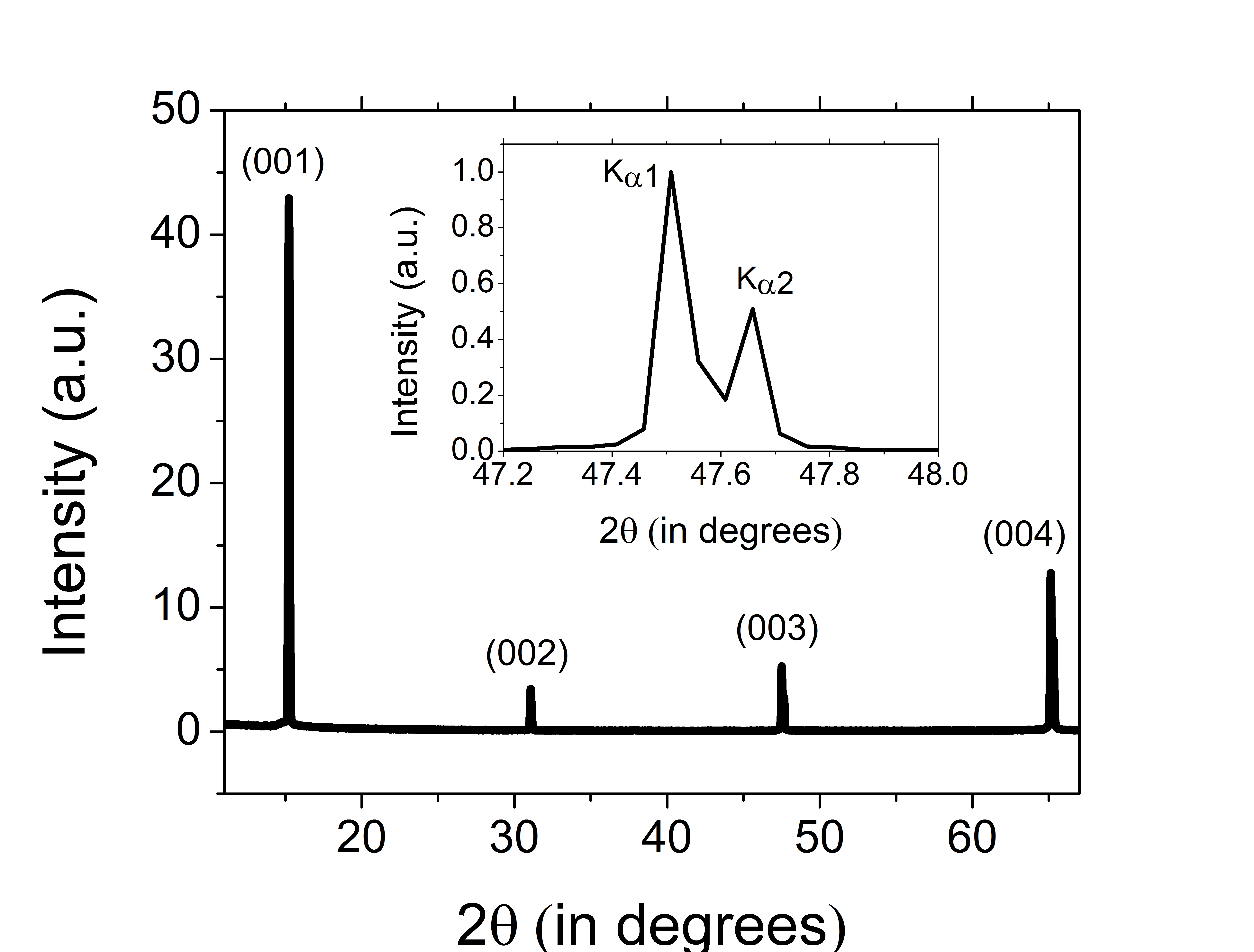}
\caption{X-Ray Diffraction spectra of pristine TiS$_2$. Inset shows the doublet peak near 2$\theta$~=~47.65$^\circ$.}
\end{figure}

\subsection{Transport measurements}

\begin{figure}[H]
\centering
\includegraphics[width=8cm]{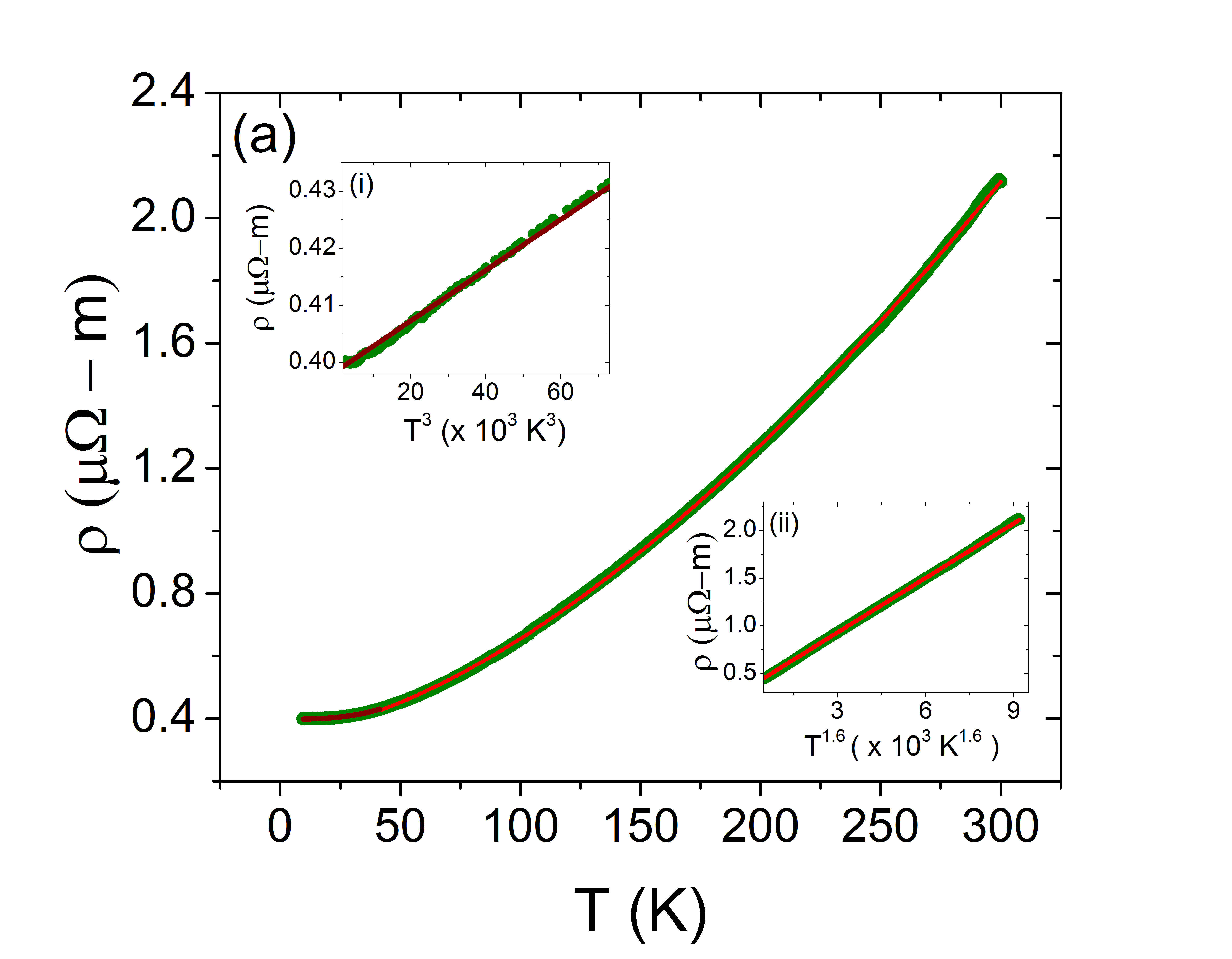}
\includegraphics[width=8cm]{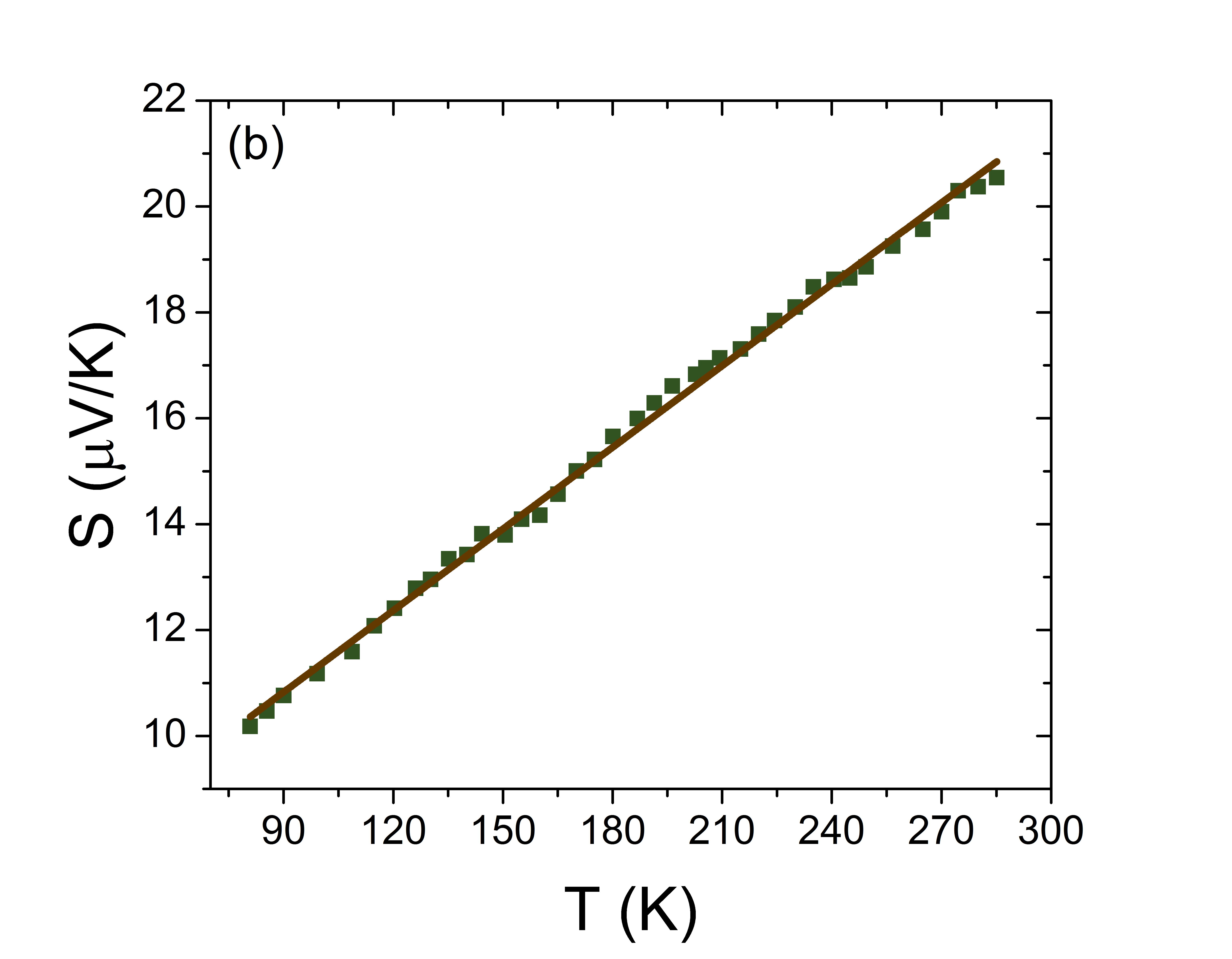}

\caption{(a) $\rho$ vs $T$ in the temperature range $9-300$~K in CIP mode, with $T^3$ (brown line) and $T^{1.6}$ (red line) fitting. Inset (i) shows $\rho$ versus $T^3$ in the temperature range $9-42$~K with linear fit (brown line), inset (ii) shows data in the temperature range $42-300$~K  versus $T^{1.6}$ with linear fit (red line). Green circles are experimental data. (b) Seebeck coefficient ($S$) vs $T$ in the temperature range $80-285$~K, with linear fit. }
\end{figure}

Electron transport behavior is studied through temperature dependent resistivity measurements. Figure 3 (a) shows resistivity versus temperature in the range $8-300$~K. In general for a metal like system, resistivity varies with $T$ as $\rho(T) = \rho_0 + AT^n$, where $n$ depends on the dominant scattering mechanism and $A$ is constant. In the range $8-42$~K, resistivity varies as $T^3$ (Fig 3 (a) inset (i)). The $n=3$ dependence (Bloch-Wilson formula) implies that the transport is dominated by s-d electron scattering at low temperatures \cite{wilson,wood,baber}. For temperatures above 42~K, we clearly observe that the resistivity follows nearly $T^2$ dependence throughout the range from $42-300$~K. The best fit curve to our experimental data is for $T^{1.6}$, with fitting parameter $\rho_0 = (371 \pm 1) \times 10^{-9} \Omega$m, $A= (192 \pm 1) \times 10^{-12} \Omega$m$K^{-1.6}$. For clarity we show $\rho$ versus $T^{1.6}$ explicitly, with a linear fit in the inset (ii) of fig. 3 (a). In case of normal metals, $T^2$ dependence is attributed to e-e scattering at low temperatures \cite{Kaveh}.  However, this cannot explain the behavior in TiS$_2$ for a wide temperature range up to room temperature \cite{Liu,Kukkonen1981, Levy}.

Through careful analysis of the Fermi surface it has been established that the reason leading to $T^2$ dependence over a broad temperature range is mainly electron-phonon coupling \cite{Zandt}. The electronic band structure of TiS$_2$ shows a multi-valley band structure \cite{Murray}, where acoustic phonons assist the transport of electrons across the valley points \cite{Klipstein} leading to $T^2$ dependence of resistivity.  Klipstein et al., discuss a detailed study of transport in TiS$_2$  \cite{Klip} where $n$ depends on stoichiometry ($\rho(T) = \rho_0 + AT^n$). $n$ can vary between 1.6 and 2.3 depending on whether the dominant scattering is inter-valley or intra-valley as a consequence of stoichiometry. Quality of samples can be estimated through residual resistivity ratio (RRR) which is indicative of deviation from stoichiometry. The RRR of our sample is $\sim$5, which is comparable with the $n \sim 1.6$  in case of Ref. 20. For samples with nearly stoichiometry composition (RRR $\sim$ 12), $n$ reported was $\sim$ 2.2 in Ref. 21.  We infer that $n$ depends on the purity of the sample, charge carrier concentration which are decisive factors on whether the mechanism is inter-valley or the intra-valley scattering.

Figure 3 (b) shows in-plane Seebeck coefficient as a function of temperature in the range $80-285$~K. The in-plane Seebeck co-efficient of TiS$_2$ varies linearly with temperature as expected for metal-like systems.  Interestingly, Seebeck co-efficient follows the metal-like behavior, though the electron conductivity does not resemble normal metal. Overall reduction in thermal conductivity, and increased electronic conductivity due to inter-valley scattering \citep{Imai} results in a significantly large Seebeck co-efficient. The magnitude of Seebeck co-efficient increases with decrease in the number of layers \cite{Zhang2012} and hence a larger Seebeck coefficient is expected for thinner flakes.

\begin{figure}[H]
\centering
\includegraphics[width=8cm]{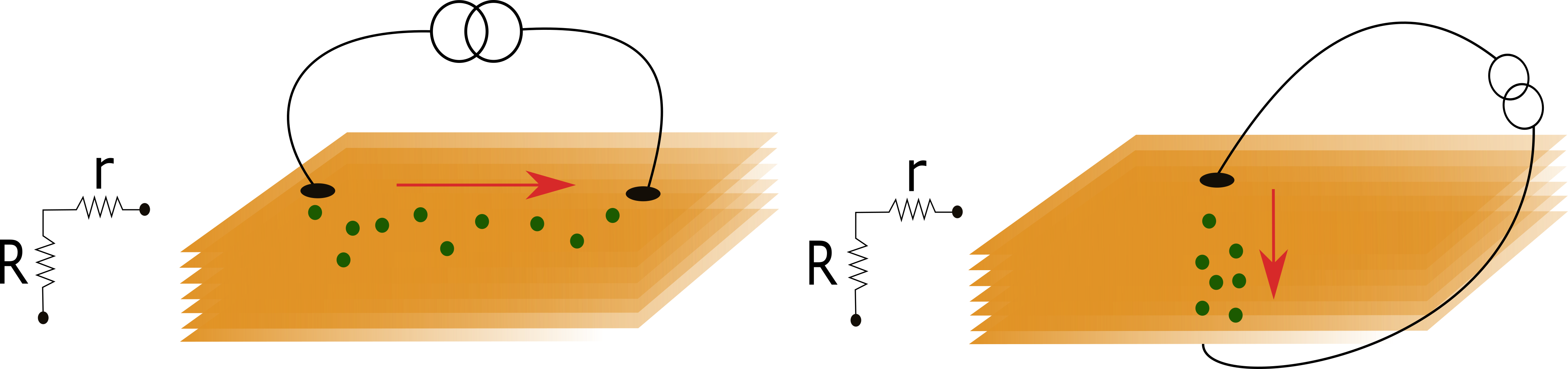}
\caption{Schematics for electron transport in the in-plane current mode (left) and perpendicular to plane current mode (right). Green solid circles represent electrons. Red arrows indicate the direction of electron flow.}
\end{figure}

TMDC possess anisotropy in structural configuration in the in-plane and perpendicular-to-plane directions due to presence of  van der Waals gap between the layers. Hence the electron transport ought to be different in the two configurations (Fig. 4).  Fig. 5 shows resistivity as a function of temperature in the CPP mode.  The magnitude of resistivity is 4 orders larger than that in CIP (R = 10$^4 \times$ r in Fig. 4), which indicates high degree of structural anisotropy. This implies that the transport in CIP mode is largely two dimensional due to the fact that electrons have a lower in-plane resistance than the inter-planar resistance, as shown in the schematic in fig. 4. However, when measurements are performed in the CPP mode the electrons are forced into conduction across layers and hence more closely represents bulk three dimensional transport.  We find that resistivity varies with temperature as $T^{1.6}$ upto 200 K and further goes as $T^5$ till 300K. The best fit equation for temperature below 200 K is $\rho(T) = \rho_0 + AT^{1.6}$, with fitting parameter $\rho_0 = (4250 \pm 4) \times ~10^{-5} \Omega$m, $A=(284 \pm 2) \times 10^{-8} \Omega$m$K^{-1.6}$. As explained for in-plane transport behavior, this may be attributed to electron - phonon coupling  \cite{Klipstein}. However, the best fit equation for temperature between 200 - 300 K is $\rho(T) = \rho_0 + BT^{5}$, with fitting parameter $\rho_0 = (510 \pm 1) \times ~10^{-4} \Omega$m, $B = (118 \pm 1) \times 10^{-16} \Omega$m$K^{-5}$. A high degree of structural anisotropy ($\sim$10$^4$), strongly suggests that the transport and coupling mechanism involved in CIP and CPP directions are not the same.

 \begin{figure}[H]
\centering
\includegraphics[width=8cm]{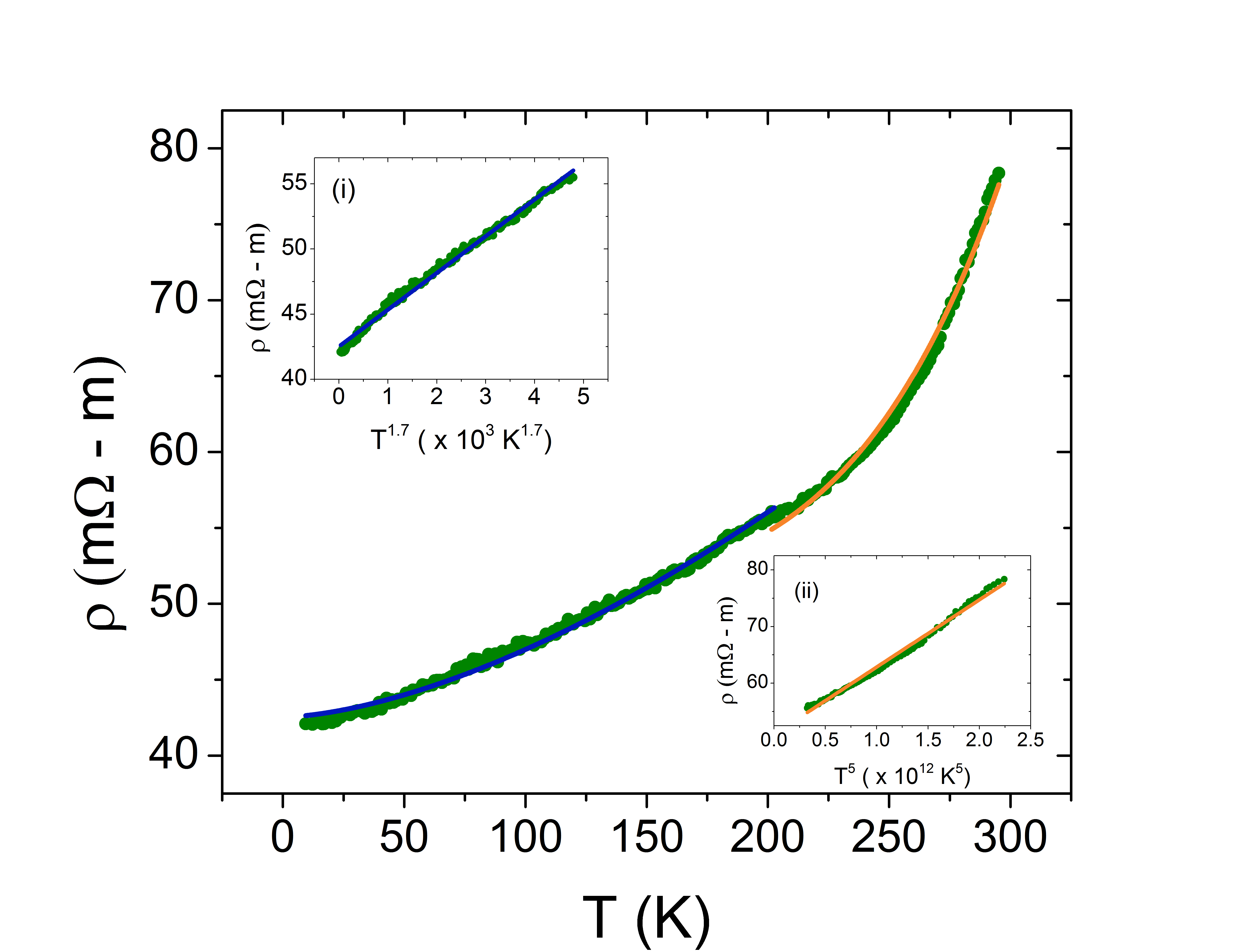}
\caption{$\rho$ vs $T$ in the temperature range $9-295$~K in CPP mode. Blue line shows $T^{1.6}$ fitting from  9 - 200 K. Orange line shows $T^{5}$ fitting from  200 - 295 K. Inset (i) shows data as a function of $T^{1.6}$ with linear fit from  9 - 200 K. Inset (ii) shows data as a function of $T^{5}$ with linear fit from   200 - 295 K. }
\end{figure}

Up to 200~K resistivity varies as T$^{1.7}$ and is attributed to electron - phonon scattering as discussed in the previous section. Interestingly, we observe an unconventional strong non-linear behavior of resistivity for temperatures above 200~K. It is well known that the temperature dependence of resistivity in normal metals follow $\rho \propto T^5$ (Bloch - Gr$\ddot{\text{u}}$neisen limit) and $\rho \propto T^3$ (Bloch - Wilson limit) in transition metals at low temperatures in the  and $\rho \propto T$ for higher temperatures \cite{Kaveh,wilson}. However, in  graphene there have been reports which demonstrate  non-linear behavior of  resistivity  at higher temperatures  \cite{Kim2016,Park2014,kim,Fratini2008}. Morozov et al. \cite{Morozov2008}  demonstrate $\rho \propto T^5$ for $T > 200~K$ and attribute it to the generation of flexural phonons. In cleaved graphene, a consequence of  corrugations leads to out-of-plane phonon modes confined in ripples above a certain temperature. Theoretically there are proposals that show existence of ripples in  MoS$_2$ above 200~K \cite{singh}. In our experiments we observe that resistivity varies as $T^{1.7}$ upto 200~K, and further varies as $T^5$ in the CPP mode. This may be attributed to generation of flexural phonons above 200~K, due to corrugations along the perpendicular-to-plane direction.  The possibility of phonon modes due to folds, bends and deformities affecting the electron transport is high in the transverse-to-plane direction compared to the in-plane direction. Hence there is no signature of $T^5$ dependence above 200~K the in-plane transport measurements.  However, a detailed theoretical model to explain this phenomenon in TiS$_2$ is yet to be explored. 

\section{Effects of irradiation} Pristine TiS$_2$ was irradiated by Au ions of energy 15 keV at normal incidence, with a fluence of $10^{16}$ ions cm$^{-2}$. XRD spectrum of the irradiated sample is shown in fig. 6. The average inter-layer spacing deduced from XRD spectrum was 5.7 \AA~ which doesn't change upon irradiation. 

\begin{figure}[H]
\centering
\includegraphics[width=9cm]{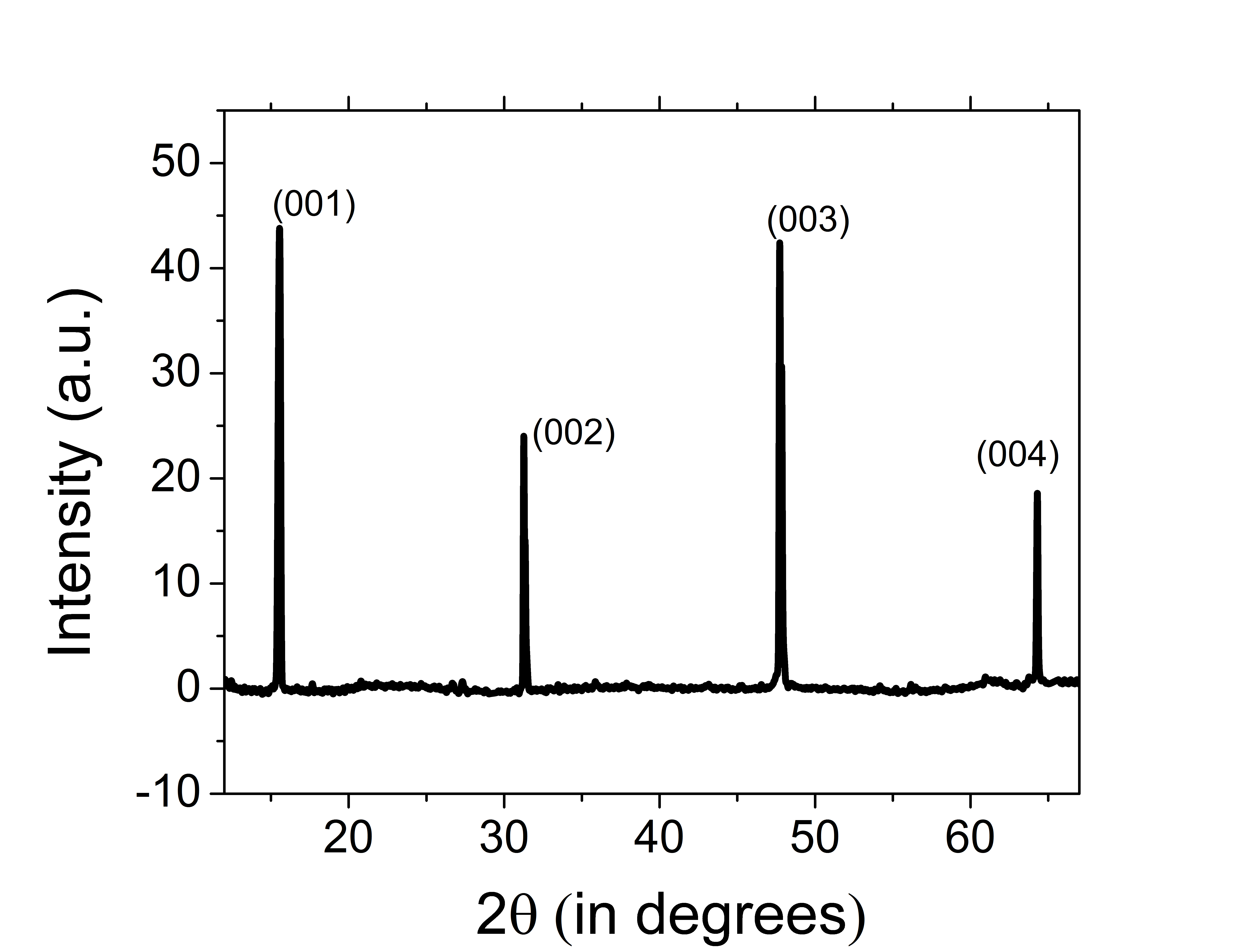}
\caption{XRD spectrum for the irradiated sample.}
\end{figure}

Figure 7 (a) and (b) shows the AFM images of the surface of TiS$_2$ before and after irradiation. The extracted profiles show an average roughness of 0.43 $\pm$ 0.02 nm in pristine and 0.24 $\pm$ 0.02  nm in case of irradiated samples.  The pristine sample has a wavy surface in this length scale. Post irradiation the induced defects flatten the surface contours. We attribute the cause of these structures and decrease in roughness to enhanced diffusion during irradiation. The evolution of these structures (nano dots) on TiS$_2$ surface might be due to self organized agglomeration of particles upon irradiation \cite{Sulania}. The typical dimensions of these dots are approximately 1 nm in height and 10-13 nm in diameter. SEM images of pristine and irradiated samples are shown in fig. 8 (a) and (b). Inset shows image of the edge. The bright lines indicate layered structure of the material. The comparison of pristine and irradiated samples show a clear evolution of nano dot-like patterns throughout the irradiated sample. This corroborates the observation of morphology made through AFM technique.

\begin{figure}[H]
\centering
\includegraphics[width=4.5cm]{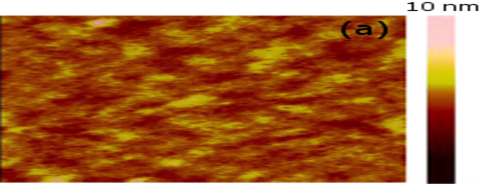}\hspace{0.7cm}
\includegraphics[width=4.5cm]{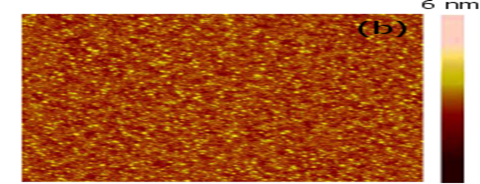}
\caption{AFM image of (a) Pristine TiS$_2$ (b) Au irradiated TiS$_2$. Area of the image: 1 $\times$ ~1 $\mu$m$^2$}
\end{figure}

\begin{figure}[H]
\centering
\includegraphics[width=4cm]{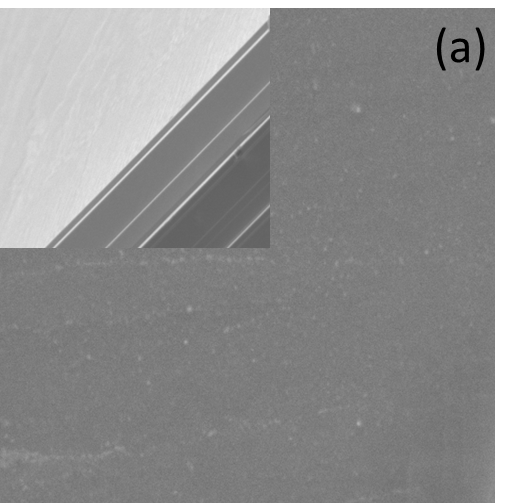}\hspace{1cm}
\includegraphics[width=4cm]{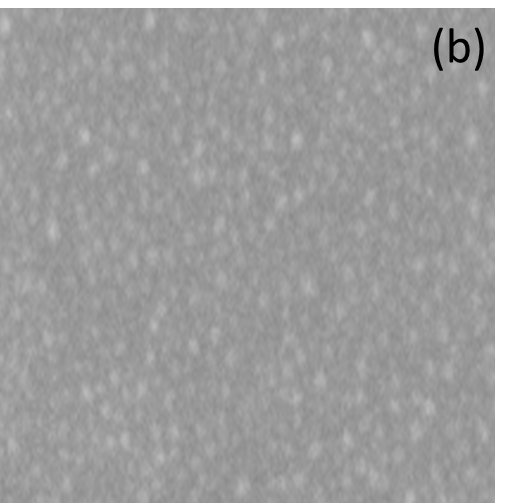}
\caption{SEM image of (a) pristine (b) irradiated TiS$_2$. Scale: 1 $\times$ 1 $\mu$m$^2$. Inset shows cross section image of pristine TiS$_2$.}
\end{figure}

To study the impact of irradiation on phonon modes we compare the Raman spectra of pristine and irradiated samples (Fig. 9). In pristine samples we observe two peaks at 230 cm$^{-1}$ and 336 cm$^{-1}$. Layered TiS$_2$  belongs to D$_{2h}$ point group symmetry and the  Raman active modes are observed at 232 cm$^{-1}$ and 336 cm$^{-1}$ which are characteristic peaks of TiS$_2$ as reported in literature \cite{Sandova,nanoscale,srep}. These correspond to Raman active modes of E$_{g}$ and A$_{1g}$ symmetry representations which are vibrational modes for in-plane and perpendicular to plane geometries, respectively. While the A$_{1g}$ mode is prominent in both pristine and irradiated samples, E$_g$  mode occurs with a much lesser intensity. This is due to strong covalent bond between the intra-layer atoms and weak van der Waals force between the layers. After irradiation the intensity of A$_{1g}$ peak is suppressed and broadened, and E$_g$ mode disappears. This implies that irradiation has destroyed the in-plane vibrational phonon modes, but the transverse modes are partially affected. The peak position of A$_{1g}$ mode at 336 cm$^{-1}$ in the pristine sample drifts towards lower wavenumber to 327 cm$^{-1}$ after irradiation.

\begin{figure}[H]
\centering
\includegraphics[width=8cm]{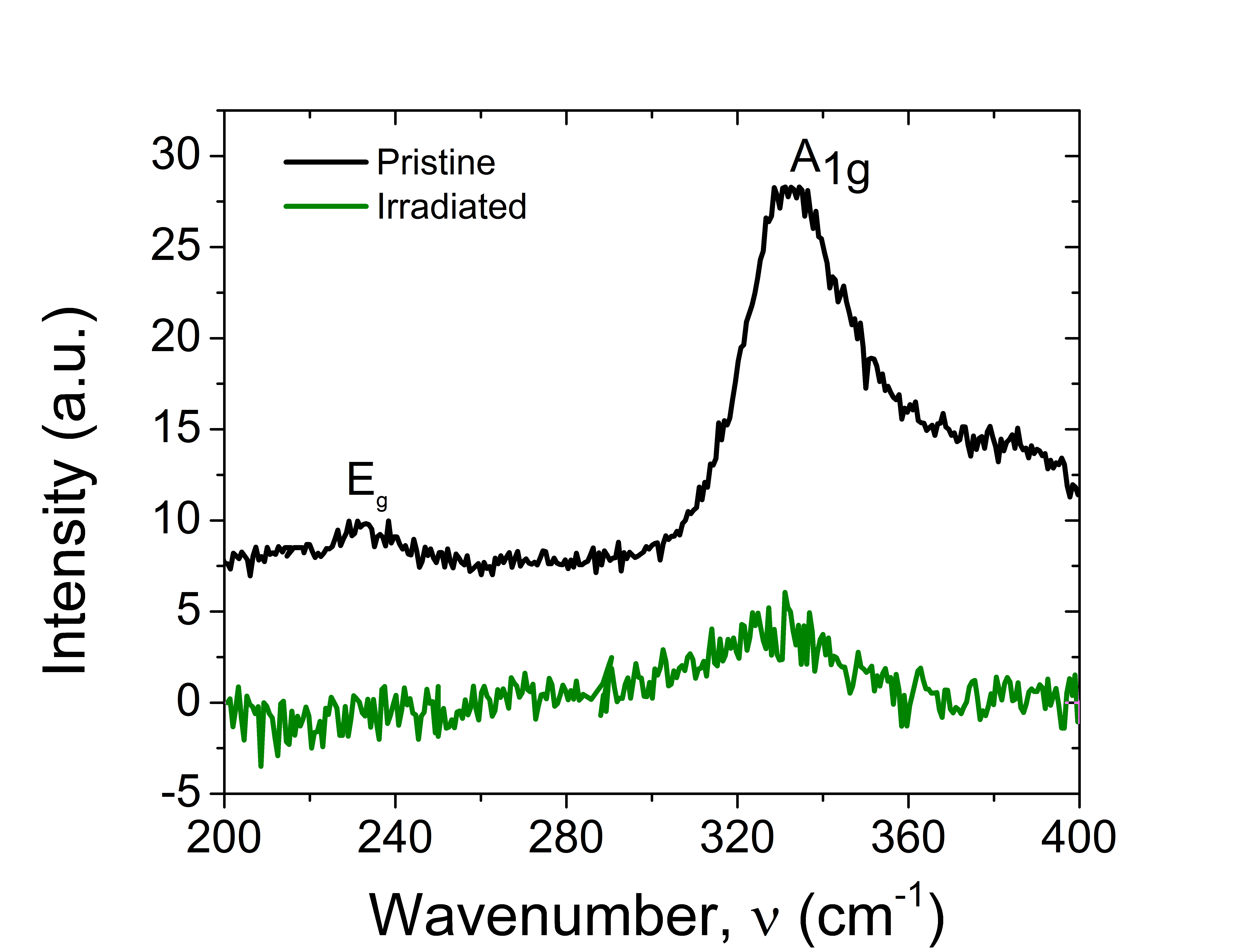}
\caption{Raman spectra for pristine and Au irradiated TiS$_2$. Black and green lines show data for pristine and irradiated samples respectively. Background baseline subtraction has been performed for irradiated data.}
\end{figure}

Figure 10 shows resistivity as a function of temperature for CIP configuration of the irradiated sample and CPP configuration of the pristine sample. Comparison of  the sample quality through RRR  shows that the irradiated  CIP has a larger RRR  compared to that in the CPP configuration which is expected. We also see that the RRR in  CIP mode of irradiated sample   is smaller than that of the the pristine CIP ($\sim$5), which indicates the damage caused to the layers. Resistivity varies in the in-plane transport mode as $T^{1.6}$ upto 200 K, and further varies as $T^5$. This is precisely the feature observed in the CPP mode transport of the pristine TiS$_2$. Generation of defects  restricts the electron transport along the plane in CIP mode and forces the electrons for conduction across the layers. Hence the transport is more like that in bulk and resembles the CPP mode measurement of pristine samples.  
\begin{figure}[H]
\centering
\includegraphics[width=8cm]{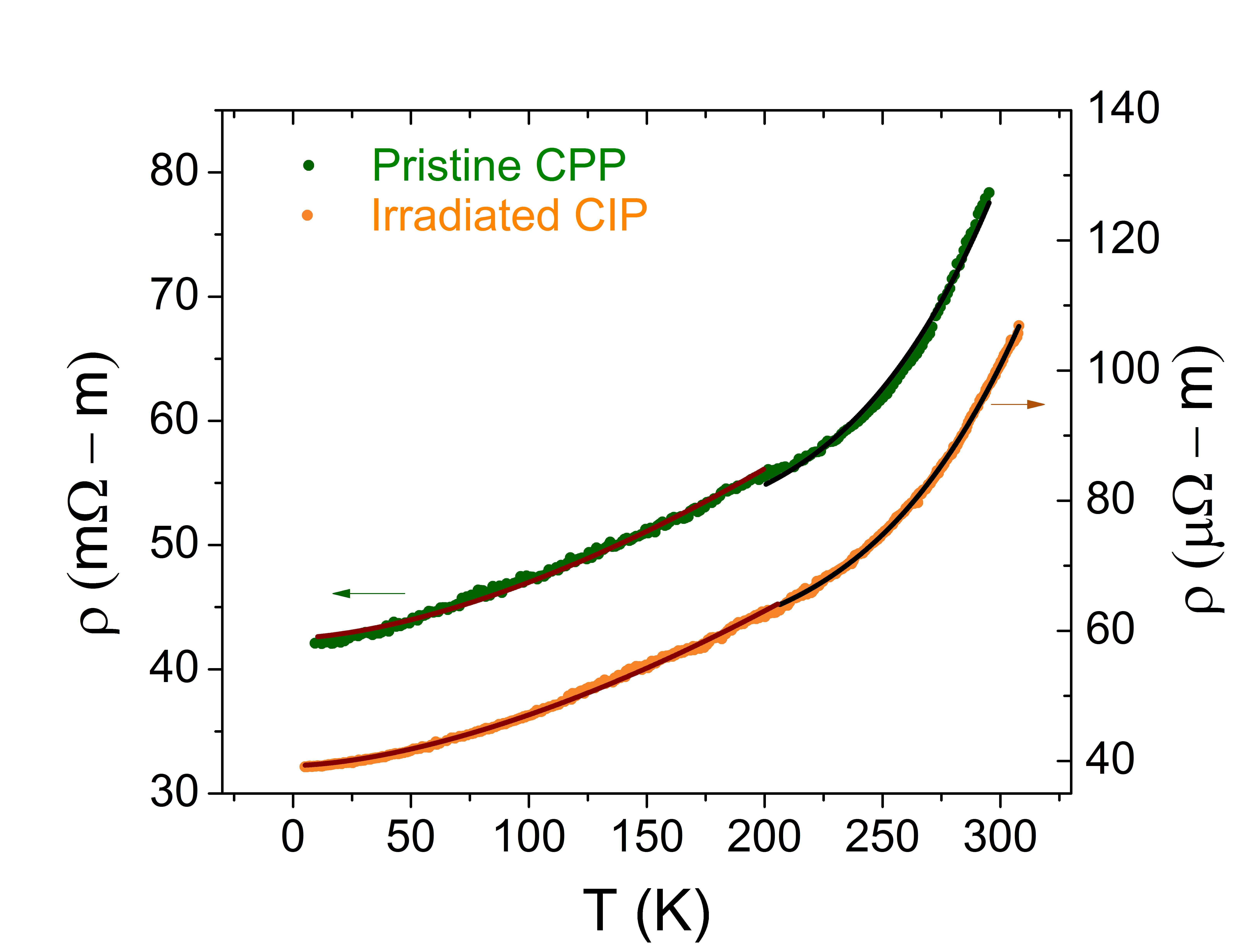}
\caption{$\rho$ versus T for Irradiated TiS$_2$ (green) in the CIP mode and pristine TiS$_2$ (orange) in the CPP mode. Solid circles indicate experimental data and lines indicate the best cit curve.}
\end{figure}

\section{Conclusions} To summarize, we have presented a detailed structural analysis and transport measurements of TiS$_2$ in its pristine and irradiated form. Resistivity varies as $T^3$ in the in-plane configuration in the temperature regime 9 - 42~K and nearly $T^2$ in the range 42 - 300~K, where as the CPP configuration shows $T^5$ dependence at higher temperatures (200 - 300 K). This feature repeats in the CIP mode transport of irradiated samples.  This maybe attributed to scattering due to flexural phonons. Seebeck co-efficient is significant at room temperature and is expected to increase with reduction in the number of layers. Hence, TiS$_2$ is a potential candidate for thermopower based device applications. X-ray diffraction shows that irradiation leads to relaxation due to strain. AFM and SEM imaging techniques establish a significant change in the surface morphology upon Au irradiation. Raman spectra on the irradiated samples show absence of E$_g$ mode and suppressed A$_{1g}$. This demonstrates the defects caused upon irradiation have significantly altered the vibrational states. In conclusion, we report a detailed structural and transport properties of TiS$_2$ and present features that are intriguing for device applications and for fundamental understanding of the transport mechanism. 

\section{Acknowledgements} Authors thank the Ion Beam Laboratory staff at Institute of Physics, Bhubaneshwar for assistance with ion beam irradiation. DS thanks Department of Science and Technology (DST), Govt. of India for Ph.D. fellowship (No. DST/INSPIRE Fellowship/2013/742) through DST-INSPIRE program. RSP, KS, PKS thank DST, Govt. of India for funds DST-SERB (No. EMR/2016/003318).


\begin{thebibliography}{42}%
\makeatletter
\providecommand \@ifxundefined [1]{%
 \@ifx{#1\undefined}
}%
\providecommand \@ifnum [1]{%
 \ifnum #1\expandafter \@firstoftwo
 \else \expandafter \@secondoftwo
 \fi
}%
\providecommand \@ifx [1]{%
 \ifx #1\expandafter \@firstoftwo
 \else \expandafter \@secondoftwo
 \fi
}%
\providecommand \natexlab [1]{#1}%
\providecommand \enquote  [1]{``#1''}%
\providecommand \bibnamefont  [1]{#1}%
\providecommand \bibfnamefont [1]{#1}%
\providecommand \citenamefont [1]{#1}%
\providecommand \href@noop [0]{\@secondoftwo}%
\providecommand \href [0]{\begingroup \@sanitize@url \@href}%
\providecommand \@href[1]{\@@startlink{#1}\@@href}%
\providecommand \@@href[1]{\endgroup#1\@@endlink}%
\providecommand \@sanitize@url [0]{\catcode `\\12\catcode `\$12\catcode
  `\&12\catcode `\#12\catcode `\^12\catcode `\_12\catcode `\%12\relax}%
\providecommand \@@startlink[1]{}%
\providecommand \@@endlink[0]{}%
\providecommand \url  [0]{\begingroup\@sanitize@url \@url }%
\providecommand \@url [1]{\endgroup\@href {#1}{\urlprefix }}%
\providecommand \urlprefix  [0]{URL }%
\providecommand \Eprint [0]{\href }%
\providecommand \doibase [0]{http://dx.doi.org/}%
\providecommand \selectlanguage [0]{\@gobble}%
\providecommand \bibinfo  [0]{\@secondoftwo}%
\providecommand \bibfield  [0]{\@secondoftwo}%
\providecommand \translation [1]{[#1]}%
\providecommand \BibitemOpen [0]{}%
\providecommand \bibitemStop [0]{}%
\providecommand \bibitemNoStop [0]{.\EOS\space}%
\providecommand \EOS [0]{\spacefactor3000\relax}%
\providecommand \BibitemShut  [1]{\csname bibitem#1\endcsname}%
\let\auto@bib@innerbib\@empty
\bibitem [{\citenamefont {Snyder}\ and\ \citenamefont
  {Toberer}(2008)}]{Snyder}%
  \BibitemOpen
  \bibfield  {author} {\bibinfo {author} {\bibfnamefont {G.~J.}\ \bibnamefont
  {Snyder}}\ and\ \bibinfo {author} {\bibfnamefont {E.~S.}\ \bibnamefont
  {Toberer}},\ }\href@noop {} {\bibfield  {journal} {\bibinfo  {journal}
  {Nature Materials}\ }\textbf {\bibinfo {volume} {7}},\ \bibinfo {pages} {105}
  (\bibinfo {year} {2008})}\BibitemShut {NoStop}%
\bibitem [{\citenamefont {Radisavljevic}\ \emph {et~al.}(2010)\citenamefont
  {Radisavljevic}, \citenamefont {Radenovic}, \citenamefont {Brivio},
  \citenamefont {Giacometti},\ and\ \citenamefont {Kis}}]{radi}%
  \BibitemOpen
  \bibfield  {author} {\bibinfo {author} {\bibfnamefont {B.}~\bibnamefont
  {Radisavljevic}}, \bibinfo {author} {\bibfnamefont {A.}~\bibnamefont
  {Radenovic}}, \bibinfo {author} {\bibfnamefont {J.}~\bibnamefont {Brivio}},
  \bibinfo {author} {\bibfnamefont {V.}~\bibnamefont {Giacometti}}, \ and\
  \bibinfo {author} {\bibfnamefont {A.}~\bibnamefont {Kis}},\ }\href {\doibase
  10.1038/nnano.2010.279} {\bibfield  {journal} {\bibinfo  {journal} {Nature
  Nanotechnology}\ }\textbf {\bibinfo {volume} {6}},\ \bibinfo {pages} {147}
  (\bibinfo {year} {2010})}\BibitemShut {NoStop}%
\bibitem [{\citenamefont {Duan}\ \emph {et~al.}(2015)\citenamefont {Duan},
  \citenamefont {Wang}, \citenamefont {Pan}, \citenamefont {Yu},\ and\
  \citenamefont {Duan}}]{rscrev}%
  \BibitemOpen
  \bibfield  {author} {\bibinfo {author} {\bibfnamefont {X.}~\bibnamefont
  {Duan}}, \bibinfo {author} {\bibfnamefont {C.}~\bibnamefont {Wang}}, \bibinfo
  {author} {\bibfnamefont {A.}~\bibnamefont {Pan}}, \bibinfo {author}
  {\bibfnamefont {R.}~\bibnamefont {Yu}}, \ and\ \bibinfo {author}
  {\bibfnamefont {X.}~\bibnamefont {Duan}},\ }\href {\doibase
  10.1039/C5CS00507H} {\bibfield  {journal} {\bibinfo  {journal} {Chemical
  Society Reviews}\ }\textbf {\bibinfo {volume} {44}},\ \bibinfo {pages} {8859}
  (\bibinfo {year} {2015})}\BibitemShut {NoStop}%
\bibitem [{\citenamefont {Jiang}\ \emph {et~al.}(2017)\citenamefont {Jiang},
  \citenamefont {Liu}, \citenamefont {Sun}, \citenamefont {Yang}, \citenamefont
  {Rajamathi}, \citenamefont {Qi}, \citenamefont {Yang}, \citenamefont {Chen},
  \citenamefont {Peng}, \citenamefont {Hwang}, \citenamefont {Sun},
  \citenamefont {Mo}, \citenamefont {Vobornik}, \citenamefont {Fujii},
  \citenamefont {Parkin}, \citenamefont {Felser}, \citenamefont {Yan},\ and\
  \citenamefont {Chen}}]{Jiang}%
  \BibitemOpen
  \bibfield  {author} {\bibinfo {author} {\bibfnamefont {J.}~\bibnamefont
  {Jiang}}, \bibinfo {author} {\bibfnamefont {Z.}~\bibnamefont {Liu}}, \bibinfo
  {author} {\bibfnamefont {Y.}~\bibnamefont {Sun}}, \bibinfo {author}
  {\bibfnamefont {H.}~\bibnamefont {Yang}}, \bibinfo {author} {\bibfnamefont
  {C.}~\bibnamefont {Rajamathi}}, \bibinfo {author} {\bibfnamefont
  {Y.}~\bibnamefont {Qi}}, \bibinfo {author} {\bibfnamefont {L.}~\bibnamefont
  {Yang}}, \bibinfo {author} {\bibfnamefont {C.}~\bibnamefont {Chen}}, \bibinfo
  {author} {\bibfnamefont {H.}~\bibnamefont {Peng}}, \bibinfo {author}
  {\bibfnamefont {C.-C.}\ \bibnamefont {Hwang}}, \bibinfo {author}
  {\bibfnamefont {S.}~\bibnamefont {Sun}}, \bibinfo {author} {\bibfnamefont
  {S.-K.}\ \bibnamefont {Mo}}, \bibinfo {author} {\bibfnamefont
  {I.}~\bibnamefont {Vobornik}}, \bibinfo {author} {\bibfnamefont
  {J.}~\bibnamefont {Fujii}}, \bibinfo {author} {\bibfnamefont
  {S.}~\bibnamefont {Parkin}}, \bibinfo {author} {\bibfnamefont
  {C.}~\bibnamefont {Felser}}, \bibinfo {author} {\bibfnamefont
  {B.}~\bibnamefont {Yan}}, \ and\ \bibinfo {author} {\bibfnamefont
  {Y.}~\bibnamefont {Chen}},\ }\href
  {http://www.nature.com/doifinder/10.1038/ncomms13973} {\bibfield  {journal}
  {\bibinfo  {journal} {Nature Communications}\ }\textbf {\bibinfo {volume}
  {8}},\ \bibinfo {pages} {13973} (\bibinfo {year} {2017})}\BibitemShut
  {NoStop}%
\bibitem [{\citenamefont {Deng}\ \emph {et~al.}(2016)\citenamefont {Deng},
  \citenamefont {Wan}, \citenamefont {Deng}, \citenamefont {Zhang},
  \citenamefont {Ding}, \citenamefont {Wang}, \citenamefont {Yan},
  \citenamefont {Huang}, \citenamefont {Zhang}, \citenamefont {Xu},
  \citenamefont {Denlinger}, \citenamefont {Fedorov}, \citenamefont {Yang},
  \citenamefont {Duan}, \citenamefont {Yao}, \citenamefont {Wu}, \citenamefont
  {Fan}, \citenamefont {Zhang}, \citenamefont {Chen},\ and\ \citenamefont
  {Zhou}}]{Deng}%
  \BibitemOpen
  \bibfield  {author} {\bibinfo {author} {\bibfnamefont {K.}~\bibnamefont
  {Deng}}, \bibinfo {author} {\bibfnamefont {G.}~\bibnamefont {Wan}}, \bibinfo
  {author} {\bibfnamefont {P.}~\bibnamefont {Deng}}, \bibinfo {author}
  {\bibfnamefont {K.}~\bibnamefont {Zhang}}, \bibinfo {author} {\bibfnamefont
  {S.}~\bibnamefont {Ding}}, \bibinfo {author} {\bibfnamefont {E.}~\bibnamefont
  {Wang}}, \bibinfo {author} {\bibfnamefont {M.}~\bibnamefont {Yan}}, \bibinfo
  {author} {\bibfnamefont {H.}~\bibnamefont {Huang}}, \bibinfo {author}
  {\bibfnamefont {H.}~\bibnamefont {Zhang}}, \bibinfo {author} {\bibfnamefont
  {Z.}~\bibnamefont {Xu}}, \bibinfo {author} {\bibfnamefont {J.}~\bibnamefont
  {Denlinger}}, \bibinfo {author} {\bibfnamefont {A.}~\bibnamefont {Fedorov}},
  \bibinfo {author} {\bibfnamefont {H.}~\bibnamefont {Yang}}, \bibinfo {author}
  {\bibfnamefont {W.}~\bibnamefont {Duan}}, \bibinfo {author} {\bibfnamefont
  {H.}~\bibnamefont {Yao}}, \bibinfo {author} {\bibfnamefont {Y.}~\bibnamefont
  {Wu}}, \bibinfo {author} {\bibfnamefont {y.~S.}\ \bibnamefont {Fan}},
  \bibinfo {author} {\bibfnamefont {H.}~\bibnamefont {Zhang}}, \bibinfo
  {author} {\bibfnamefont {X.}~\bibnamefont {Chen}}, \ and\ \bibinfo {author}
  {\bibfnamefont {S.}~\bibnamefont {Zhou}},\ }\href {\doibase
  10.1038/nphys3871} {\bibfield  {journal} {\bibinfo  {journal} {Nature
  Physics}\ }\textbf {\bibinfo {volume} {1}},\ \bibinfo {pages} {1} (\bibinfo
  {year} {2016})},\ \Eprint {http://arxiv.org/abs/1603.08508} {1603.08508}
  \BibitemShut {NoStop}%
\bibitem [{\citenamefont {Soluyanov}\ \emph {et~al.}(2015)\citenamefont
  {Soluyanov}, \citenamefont {Gresch}, \citenamefont {Wang}, \citenamefont
  {Wu}, \citenamefont {Troyer}, \citenamefont {Dai},\ and\ \citenamefont
  {Bernevig}}]{Soluyanov}%
  \BibitemOpen
  \bibfield  {author} {\bibinfo {author} {\bibfnamefont {A.~A.}\ \bibnamefont
  {Soluyanov}}, \bibinfo {author} {\bibfnamefont {D.}~\bibnamefont {Gresch}},
  \bibinfo {author} {\bibfnamefont {Z.}~\bibnamefont {Wang}}, \bibinfo {author}
  {\bibfnamefont {Q.}~\bibnamefont {Wu}}, \bibinfo {author} {\bibfnamefont
  {M.}~\bibnamefont {Troyer}}, \bibinfo {author} {\bibfnamefont
  {X.}~\bibnamefont {Dai}}, \ and\ \bibinfo {author} {\bibfnamefont {B.~A.}\
  \bibnamefont {Bernevig}},\ }\href@noop {} {\bibfield  {journal} {\bibinfo
  {journal} {Nature}\ }\textbf {\bibinfo {volume} {527}},\ \bibinfo {pages}
  {495} (\bibinfo {year} {2015})}\BibitemShut {NoStop}%
\bibitem [{\citenamefont {Lu}\ \emph {et~al.}(2016)\citenamefont {Lu},
  \citenamefont {Kim}, \citenamefont {Yang}, \citenamefont {Gao}, \citenamefont
  {Wu}, \citenamefont {Shao}, \citenamefont {Li}, \citenamefont {Zhou},
  \citenamefont {Sun}, \citenamefont {Akinwande}, \citenamefont {Xing},\ and\
  \citenamefont {Lin}}]{Lu2016}%
  \BibitemOpen
  \bibfield  {author} {\bibinfo {author} {\bibfnamefont {P.}~\bibnamefont
  {Lu}}, \bibinfo {author} {\bibfnamefont {J.-S.}\ \bibnamefont {Kim}},
  \bibinfo {author} {\bibfnamefont {J.}~\bibnamefont {Yang}}, \bibinfo {author}
  {\bibfnamefont {H.}~\bibnamefont {Gao}}, \bibinfo {author} {\bibfnamefont
  {J.}~\bibnamefont {Wu}}, \bibinfo {author} {\bibfnamefont {D.}~\bibnamefont
  {Shao}}, \bibinfo {author} {\bibfnamefont {B.}~\bibnamefont {Li}}, \bibinfo
  {author} {\bibfnamefont {D.}~\bibnamefont {Zhou}}, \bibinfo {author}
  {\bibfnamefont {J.}~\bibnamefont {Sun}}, \bibinfo {author} {\bibfnamefont
  {D.}~\bibnamefont {Akinwande}}, \bibinfo {author} {\bibfnamefont
  {D.}~\bibnamefont {Xing}}, \ and\ \bibinfo {author} {\bibfnamefont {J.-F.}\
  \bibnamefont {Lin}},\ }\href@noop {} {\bibfield  {journal} {\bibinfo
  {journal} {Physical Review B}\ }\textbf {\bibinfo {volume} {94}},\ \bibinfo
  {pages} {224512} (\bibinfo {year} {2016})}\BibitemShut {NoStop}%
\bibitem [{\citenamefont {Avsar}\ \emph {et~al.}(2014)\citenamefont {Avsar},
  \citenamefont {Tan}, \citenamefont {Taychatanapat}, \citenamefont
  {Balakrishnan}, \citenamefont {Koon}, \citenamefont {Yeo}, \citenamefont
  {Lahiri}, \citenamefont {Carvalho}, \citenamefont {Rodin}, \citenamefont
  {O'Farrell}, \citenamefont {Eda}, \citenamefont {{Castro Neto}},\ and\
  \citenamefont {{\"{O}}zyilmaz}}]{Avsar2014}%
  \BibitemOpen
  \bibfield  {author} {\bibinfo {author} {\bibfnamefont {A.}~\bibnamefont
  {Avsar}}, \bibinfo {author} {\bibfnamefont {J.~Y.}\ \bibnamefont {Tan}},
  \bibinfo {author} {\bibfnamefont {T.}~\bibnamefont {Taychatanapat}}, \bibinfo
  {author} {\bibfnamefont {J.}~\bibnamefont {Balakrishnan}}, \bibinfo {author}
  {\bibfnamefont {G.}~\bibnamefont {Koon}}, \bibinfo {author} {\bibfnamefont
  {Y.}~\bibnamefont {Yeo}}, \bibinfo {author} {\bibfnamefont {J.}~\bibnamefont
  {Lahiri}}, \bibinfo {author} {\bibfnamefont {A.}~\bibnamefont {Carvalho}},
  \bibinfo {author} {\bibfnamefont {A.~S.}\ \bibnamefont {Rodin}}, \bibinfo
  {author} {\bibfnamefont {E.}~\bibnamefont {O'Farrell}}, \bibinfo {author}
  {\bibfnamefont {G.}~\bibnamefont {Eda}}, \bibinfo {author} {\bibfnamefont
  {A.~H.}\ \bibnamefont {{Castro Neto}}}, \ and\ \bibinfo {author}
  {\bibfnamefont {B.}~\bibnamefont {{\"{O}}zyilmaz}},\ }\href {\doibase
  10.1038/ncomms5875} {\bibfield  {journal} {\bibinfo  {journal} {Nature
  Communications}\ }\textbf {\bibinfo {volume} {5}},\ \bibinfo {pages} {4875}
  (\bibinfo {year} {2014})}\BibitemShut {NoStop}%
\bibitem [{\citenamefont {Lee}\ \emph {et~al.}(2013)\citenamefont {Lee},
  \citenamefont {Yu}, \citenamefont {Cui}, \citenamefont {Petrone},
  \citenamefont {Lee}, \citenamefont {Choi}, \citenamefont {Lee}, \citenamefont
  {Lee}, \citenamefont {Yoo}, \citenamefont {Watanabe}, \citenamefont
  {Taniguchi}, \citenamefont {Nuckolls}, \citenamefont {Kim},\ and\
  \citenamefont {Hone}}]{Lee2013}%
  \BibitemOpen
  \bibfield  {author} {\bibinfo {author} {\bibfnamefont {G.-H.}\ \bibnamefont
  {Lee}}, \bibinfo {author} {\bibfnamefont {Y.-J.}\ \bibnamefont {Yu}},
  \bibinfo {author} {\bibfnamefont {X.}~\bibnamefont {Cui}}, \bibinfo {author}
  {\bibfnamefont {N.}~\bibnamefont {Petrone}}, \bibinfo {author} {\bibfnamefont
  {C.-H.}\ \bibnamefont {Lee}}, \bibinfo {author} {\bibfnamefont {M.~S.}\
  \bibnamefont {Choi}}, \bibinfo {author} {\bibfnamefont {D.-Y.}\ \bibnamefont
  {Lee}}, \bibinfo {author} {\bibfnamefont {C.}~\bibnamefont {Lee}}, \bibinfo
  {author} {\bibfnamefont {W.~J.}\ \bibnamefont {Yoo}}, \bibinfo {author}
  {\bibfnamefont {K.}~\bibnamefont {Watanabe}}, \bibinfo {author}
  {\bibfnamefont {T.}~\bibnamefont {Taniguchi}}, \bibinfo {author}
  {\bibfnamefont {C.}~\bibnamefont {Nuckolls}}, \bibinfo {author}
  {\bibfnamefont {P.}~\bibnamefont {Kim}}, \ and\ \bibinfo {author}
  {\bibfnamefont {J.}~\bibnamefont {Hone}},\ }\href@noop {} {\bibfield
  {journal} {\bibinfo  {journal} {ACS Nano}\ }\textbf {\bibinfo {volume} {7}},\
  \bibinfo {pages} {7931} (\bibinfo {year} {2013})}\BibitemShut {NoStop}%
\bibitem [{\citenamefont {Langouche}, \citenamefont {Kamalakar},\ and\
  \citenamefont {Dash}(2013)}]{Langouche2013}%
  \BibitemOpen
  \bibfield  {author} {\bibinfo {author} {\bibfnamefont {L.}~\bibnamefont
  {Langouche}}, \bibinfo {author} {\bibfnamefont {M.~V.}\ \bibnamefont
  {Kamalakar}}, \ and\ \bibinfo {author} {\bibfnamefont {S.~P.}\ \bibnamefont
  {Dash}},\ }\href {http://pubs.acs.org/doi/abs/10.1021/nn402954e} {\bibfield
  {journal} {\bibinfo  {journal} {ACS Nano}\ }\textbf {\bibinfo {volume} {7}},\
  \bibinfo {pages} {7931} (\bibinfo {year} {2013})}\BibitemShut {NoStop}%
\bibitem [{\citenamefont {Larentis}, \citenamefont {Fallahazad},\ and\
  \citenamefont {Tutuc}(2012)}]{Larentis}%
  \BibitemOpen
  \bibfield  {author} {\bibinfo {author} {\bibfnamefont {S.}~\bibnamefont
  {Larentis}}, \bibinfo {author} {\bibfnamefont {B.}~\bibnamefont
  {Fallahazad}}, \ and\ \bibinfo {author} {\bibfnamefont {E.}~\bibnamefont
  {Tutuc}},\ }\href@noop {} {\bibfield  {journal} {\bibinfo  {journal} {Applied
  Physics Letters}\ }\textbf {\bibinfo {volume} {101}},\ \bibinfo {pages}
  {223104} (\bibinfo {year} {2012})}\BibitemShut {NoStop}%
\bibitem [{\citenamefont {Ali}\ \emph {et~al.}(2014)\citenamefont {Ali},
  \citenamefont {Xiong}, \citenamefont {Flynn}, \citenamefont {Tao},
  \citenamefont {Gibson}, \citenamefont {Schoop}, \citenamefont {Liang},
  \citenamefont {Haldolaarachchige}, \citenamefont {Hirschberger},
  \citenamefont {Ong},\ and\ \citenamefont {Cava}}]{Ali2014}%
  \BibitemOpen
  \bibfield  {author} {\bibinfo {author} {\bibfnamefont {M.~N.}\ \bibnamefont
  {Ali}}, \bibinfo {author} {\bibfnamefont {J.}~\bibnamefont {Xiong}}, \bibinfo
  {author} {\bibfnamefont {S.}~\bibnamefont {Flynn}}, \bibinfo {author}
  {\bibfnamefont {J.}~\bibnamefont {Tao}}, \bibinfo {author} {\bibfnamefont
  {Q.~D.}\ \bibnamefont {Gibson}}, \bibinfo {author} {\bibfnamefont {L.~M.}\
  \bibnamefont {Schoop}}, \bibinfo {author} {\bibfnamefont {T.}~\bibnamefont
  {Liang}}, \bibinfo {author} {\bibfnamefont {N.}~\bibnamefont
  {Haldolaarachchige}}, \bibinfo {author} {\bibfnamefont {M.}~\bibnamefont
  {Hirschberger}}, \bibinfo {author} {\bibfnamefont {N.~P.}\ \bibnamefont
  {Ong}}, \ and\ \bibinfo {author} {\bibfnamefont {R.~J.}\ \bibnamefont
  {Cava}},\ }\href {\doibase 10.1038/nature13763} {\bibfield  {journal}
  {\bibinfo  {journal} {Nature}\ }\textbf {\bibinfo {volume} {514}},\ \bibinfo
  {pages} {205} (\bibinfo {year} {2014})},\ \Eprint
  {http://arxiv.org/abs/1405.0973} {arXiv:1405.0973} \BibitemShut {NoStop}%
\bibitem [{\citenamefont {Keum}\ \emph {et~al.}(2015)\citenamefont {Keum},
  \citenamefont {Cho}, \citenamefont {Kim}, \citenamefont {Choe}, \citenamefont
  {Sung}, \citenamefont {Kan}, \citenamefont {Kang}, \citenamefont {Hwang},
  \citenamefont {Kim}, \citenamefont {Yang}, \citenamefont {Chang},\ and\
  \citenamefont {Hee}}]{Keum2015}%
  \BibitemOpen
  \bibfield  {author} {\bibinfo {author} {\bibfnamefont {D.~H.}\ \bibnamefont
  {Keum}}, \bibinfo {author} {\bibfnamefont {S.}~\bibnamefont {Cho}}, \bibinfo
  {author} {\bibfnamefont {J.~H.}\ \bibnamefont {Kim}}, \bibinfo {author}
  {\bibfnamefont {D.-H.}\ \bibnamefont {Choe}}, \bibinfo {author}
  {\bibfnamefont {H.-J.}\ \bibnamefont {Sung}}, \bibinfo {author}
  {\bibfnamefont {M.}~\bibnamefont {Kan}}, \bibinfo {author} {\bibfnamefont
  {H.}~\bibnamefont {Kang}}, \bibinfo {author} {\bibfnamefont {J.-Y.}\
  \bibnamefont {Hwang}}, \bibinfo {author} {\bibfnamefont {S.~W.}\ \bibnamefont
  {Kim}}, \bibinfo {author} {\bibfnamefont {H.}~\bibnamefont {Yang}}, \bibinfo
  {author} {\bibfnamefont {K.~J.}\ \bibnamefont {Chang}}, \ and\ \bibinfo
  {author} {\bibfnamefont {Y.}~\bibnamefont {Hee}},\ }\href {\doibase
  10.1038/nphys3314} {\bibfield  {journal} {\bibinfo  {journal} {Nature
  Physics}\ }\textbf {\bibinfo {volume} {11}},\ \bibinfo {pages} {482}
  (\bibinfo {year} {2015})}\BibitemShut {NoStop}%
\bibitem [{\citenamefont {Mak}\ and\ \citenamefont {Shan}(2016)}]{Mak2016}%
  \BibitemOpen
  \bibfield  {author} {\bibinfo {author} {\bibfnamefont {K.~F.}\ \bibnamefont
  {Mak}}\ and\ \bibinfo {author} {\bibfnamefont {J.}~\bibnamefont {Shan}},\
  }\href {http://dx.doi.org/10.1038/nphoton.2015.282} {\bibfield  {journal}
  {\bibinfo  {journal} {Nature Photonics}\ }\textbf {\bibinfo {volume} {10}},\
  \bibinfo {pages} {216} (\bibinfo {year} {2016})}\BibitemShut {NoStop}%
\bibitem [{\citenamefont {Fang}, \citenamefont {de~Groot},\ and\ \citenamefont
  {Haas}(1997)}]{Fang1997}%
  \BibitemOpen
  \bibfield  {author} {\bibinfo {author} {\bibfnamefont {C.~M.}\ \bibnamefont
  {Fang}}, \bibinfo {author} {\bibfnamefont {R.~A.}\ \bibnamefont {de~Groot}},
  \ and\ \bibinfo {author} {\bibfnamefont {C.}~\bibnamefont {Haas}},\
  }\href@noop {} {\bibfield  {journal} {\bibinfo  {journal} {Physical Review
  B}\ }\textbf {\bibinfo {volume} {56}},\ \bibinfo {pages} {4455} (\bibinfo
  {year} {1997})}\BibitemShut {NoStop}%
\bibitem [{\citenamefont {Julien}\ \emph {et~al.}(2016)\citenamefont {Julien},
  \citenamefont {Mauger}, \citenamefont {Vijh},\ and\ \citenamefont
  {Zaghib}}]{battery}%
  \BibitemOpen
  \bibfield  {author} {\bibinfo {author} {\bibfnamefont {C.}~\bibnamefont
  {Julien}}, \bibinfo {author} {\bibfnamefont {A.}~\bibnamefont {Mauger}},
  \bibinfo {author} {\bibfnamefont {A.}~\bibnamefont {Vijh}}, \ and\ \bibinfo
  {author} {\bibfnamefont {K.}~\bibnamefont {Zaghib}},\ }in\ \href@noop {}
  {\emph {\bibinfo {booktitle} {Lithium Batteries: Science and Technology}}}\
  (\bibinfo  {publisher} {Springer},\ \bibinfo {year} {2016})\BibitemShut
  {NoStop}%
\bibitem [{\citenamefont {Klipstein}\ and\ \citenamefont
  {Friend}(1984)}]{Klipstein}%
  \BibitemOpen
  \bibfield  {author} {\bibinfo {author} {\bibfnamefont {P.~C.}\ \bibnamefont
  {Klipstein}}\ and\ \bibinfo {author} {\bibfnamefont {R.~H.}\ \bibnamefont
  {Friend}},\ }\href@noop {} {\bibfield  {journal} {\bibinfo  {journal}
  {Journal of Physics C: Solid State Physics}\ }\textbf {\bibinfo {volume}
  {17}},\ \bibinfo {pages} {2713} (\bibinfo {year} {1984})}\BibitemShut
  {NoStop}%
\bibitem [{\citenamefont {Yu}, \citenamefont {Sun},\ and\ \citenamefont
  {Zhou}(2010)}]{Yu}%
  \BibitemOpen
  \bibfield  {author} {\bibinfo {author} {\bibfnamefont {F.}~\bibnamefont
  {Yu}}, \bibinfo {author} {\bibfnamefont {J.-X.}\ \bibnamefont {Sun}}, \ and\
  \bibinfo {author} {\bibfnamefont {Y.-H.}\ \bibnamefont {Zhou}},\ }\href
  {\doibase http://dx.doi.org/10.1016/j.solidstatesciences.2010.07.031}
  {\bibfield  {journal} {\bibinfo  {journal} {Solid State Sciences}\ }\textbf
  {\bibinfo {volume} {12}},\ \bibinfo {pages} {1786 } (\bibinfo {year}
  {2010})}\BibitemShut {NoStop}%
\bibitem [{\citenamefont {Yu}\ and\ \citenamefont {Ross}(2011)}]{ross}%
  \BibitemOpen
  \bibfield  {author} {\bibinfo {author} {\bibfnamefont {Y.~G.}\ \bibnamefont
  {Yu}}\ and\ \bibinfo {author} {\bibfnamefont {N.~L.}\ \bibnamefont {Ross}},\
  }\href {http://stacks.iop.org/0953-8984/23/i=5/a=055401} {\bibfield
  {journal} {\bibinfo  {journal} {Journal of Physics: Condensed Matter}\
  }\textbf {\bibinfo {volume} {23}},\ \bibinfo {pages} {055401} (\bibinfo
  {year} {2011})}\BibitemShut {NoStop}%
\bibitem [{\citenamefont {Klipstein}\ \emph {et~al.}(1981)\citenamefont
  {Klipstein}, \citenamefont {Bagnall}, \citenamefont {Liang}, \citenamefont
  {Marseglia},\ and\ \citenamefont {Friend}}]{Klip}%
  \BibitemOpen
  \bibfield  {author} {\bibinfo {author} {\bibfnamefont {P.~C.}\ \bibnamefont
  {Klipstein}}, \bibinfo {author} {\bibfnamefont {a.~G.}\ \bibnamefont
  {Bagnall}}, \bibinfo {author} {\bibfnamefont {W.~Y.}\ \bibnamefont {Liang}},
  \bibinfo {author} {\bibfnamefont {E.~a.}\ \bibnamefont {Marseglia}}, \ and\
  \bibinfo {author} {\bibfnamefont {R.~H.}\ \bibnamefont {Friend}},\
  }\href@noop {} {\bibfield  {journal} {\bibinfo  {journal} {Journal of Physics
  C: Solid State Physics}\ }\textbf {\bibinfo {volume} {14}},\ \bibinfo {pages}
  {4067} (\bibinfo {year} {1981})}\BibitemShut {NoStop}%
\bibitem [{\citenamefont {Imai}, \citenamefont {Shimakawa},\ and\ \citenamefont
  {Kubo}(2001)}]{Imai}%
  \BibitemOpen
  \bibfield  {author} {\bibinfo {author} {\bibfnamefont {H.}~\bibnamefont
  {Imai}}, \bibinfo {author} {\bibfnamefont {Y.}~\bibnamefont {Shimakawa}}, \
  and\ \bibinfo {author} {\bibfnamefont {Y.}~\bibnamefont {Kubo}},\ }\href
  {\doibase 10.1103/PhysRevB.64.241104} {\bibfield  {journal} {\bibinfo
  {journal} {Physical Review B}\ }\textbf {\bibinfo {volume} {64}},\ \bibinfo
  {pages} {241104} (\bibinfo {year} {2001})}\BibitemShut {NoStop}%
\bibitem [{\citenamefont {Dolui}\ and\ \citenamefont {Sanvito}()}]{Dolui}%
  \BibitemOpen
  \bibfield  {author} {\bibinfo {author} {\bibfnamefont {K.}~\bibnamefont
  {Dolui}}\ and\ \bibinfo {author} {\bibfnamefont {S.}~\bibnamefont
  {Sanvito}},\ }\href@noop {} {\bibfield  {journal} {\bibinfo  {journal}
  {Europhysics Letters}\ ,\ \bibinfo {pages} {47001}}}\Eprint
  {http://arxiv.org/abs/1310.1866} {1310.1866} \BibitemShut {NoStop}%
\bibitem [{\citenamefont {Wilson}(1938)}]{wilson}%
  \BibitemOpen
  \bibfield  {author} {\bibinfo {author} {\bibfnamefont {A.~H.}\ \bibnamefont
  {Wilson}},\ }\href@noop {} {\bibfield  {journal} {\bibinfo  {journal} {Proc.
  Roy. Soc.}\ }\textbf {\bibinfo {volume} {A167}},\ \bibinfo {pages} {580}
  (\bibinfo {year} {1938})}\BibitemShut {NoStop}%
\bibitem [{\citenamefont {White}\ and\ \citenamefont {Woods}(1959)}]{wood}%
  \BibitemOpen
  \bibfield  {author} {\bibinfo {author} {\bibfnamefont {G.~K.}\ \bibnamefont
  {White}}\ and\ \bibinfo {author} {\bibfnamefont {S.~B.}\ \bibnamefont
  {Woods}},\ }\href@noop {} {\bibfield  {journal} {\bibinfo  {journal}
  {Phiosophical Transactions of the Royal Society of London}\ }\textbf
  {\bibinfo {volume} {251}},\ \bibinfo {pages} {273} (\bibinfo {year}
  {1959})}\BibitemShut {NoStop}%
\bibitem [{\citenamefont {Baber}(1937)}]{baber}%
  \BibitemOpen
  \bibfield  {author} {\bibinfo {author} {\bibfnamefont {W.~G.}\ \bibnamefont
  {Baber}},\ }\href@noop {} {\bibfield  {journal} {\bibinfo  {journal} {Proc.
  Roy. Soc.}\ }\textbf {\bibinfo {volume} {158A}},\ \bibinfo {pages} {383}
  (\bibinfo {year} {1937})}\BibitemShut {NoStop}%
\bibitem [{\citenamefont {Kaveh}\ and\ \citenamefont {Wiser}(1984)}]{Kaveh}%
  \BibitemOpen
  \bibfield  {author} {\bibinfo {author} {\bibfnamefont {M.}~\bibnamefont
  {Kaveh}}\ and\ \bibinfo {author} {\bibfnamefont {N.}~\bibnamefont {Wiser}},\
  }\href@noop {} {\bibfield  {journal} {\bibinfo  {journal} {Advances in
  Physics}\ }\textbf {\bibinfo {volume} {33}},\ \bibinfo {pages} {257}
  (\bibinfo {year} {1984})}\BibitemShut {NoStop}%
\bibitem [{\citenamefont {Liu}\ \emph {et~al.}(2011)\citenamefont {Liu},
  \citenamefont {Yang}, \citenamefont {Han}, \citenamefont {Hu}, \citenamefont
  {Ren}, \citenamefont {Liu}, \citenamefont {Ma},\ and\ \citenamefont
  {Gao}}]{Liu}%
  \BibitemOpen
  \bibfield  {author} {\bibinfo {author} {\bibfnamefont {B.}~\bibnamefont
  {Liu}}, \bibinfo {author} {\bibfnamefont {J.}~\bibnamefont {Yang}}, \bibinfo
  {author} {\bibfnamefont {Y.}~\bibnamefont {Han}}, \bibinfo {author}
  {\bibfnamefont {T.}~\bibnamefont {Hu}}, \bibinfo {author} {\bibfnamefont
  {W.}~\bibnamefont {Ren}}, \bibinfo {author} {\bibfnamefont {C.}~\bibnamefont
  {Liu}}, \bibinfo {author} {\bibfnamefont {Y.}~\bibnamefont {Ma}}, \ and\
  \bibinfo {author} {\bibfnamefont {C.}~\bibnamefont {Gao}},\ }\href@noop {}
  {\bibfield  {journal} {\bibinfo  {journal} {Journal of Applied Physics}\
  }\textbf {\bibinfo {volume} {109}},\ \bibinfo {pages} {053717} (\bibinfo
  {year} {2011})}\BibitemShut {NoStop}%
\bibitem [{\citenamefont {Kukkonen}\ \emph {et~al.}(1981)\citenamefont
  {Kukkonen}, \citenamefont {Kaiser}, \citenamefont {Logothetis}, \citenamefont
  {Blumenstock}, \citenamefont {Schroeder}, \citenamefont {Faile},
  \citenamefont {Colella},\ and\ \citenamefont {Gambold}}]{Kukkonen1981}%
  \BibitemOpen
  \bibfield  {author} {\bibinfo {author} {\bibfnamefont {C.~A.}\ \bibnamefont
  {Kukkonen}}, \bibinfo {author} {\bibfnamefont {W.~J.}\ \bibnamefont
  {Kaiser}}, \bibinfo {author} {\bibfnamefont {E.~M.}\ \bibnamefont
  {Logothetis}}, \bibinfo {author} {\bibfnamefont {B.~J.}\ \bibnamefont
  {Blumenstock}}, \bibinfo {author} {\bibfnamefont {P.~A.}\ \bibnamefont
  {Schroeder}}, \bibinfo {author} {\bibfnamefont {S.~P.}\ \bibnamefont
  {Faile}}, \bibinfo {author} {\bibfnamefont {R.}~\bibnamefont {Colella}}, \
  and\ \bibinfo {author} {\bibfnamefont {J.}~\bibnamefont {Gambold}},\
  }\href@noop {} {\bibfield  {journal} {\bibinfo  {journal} {Physical Review
  B}\ }\textbf {\bibinfo {volume} {24}},\ \bibinfo {pages} {1691} (\bibinfo
  {year} {1981})}\BibitemShut {NoStop}%
\bibitem [{\citenamefont {Scharli}\ and\ \citenamefont {Levy}(1986)}]{Levy}%
  \BibitemOpen
  \bibfield  {author} {\bibinfo {author} {\bibfnamefont {M.}~\bibnamefont
  {Scharli}}\ and\ \bibinfo {author} {\bibfnamefont {F.}~\bibnamefont {Levy}},\
  }\href@noop {} {\bibfield  {journal} {\bibinfo  {journal} {Physical Review
  B}\ }\textbf {\bibinfo {volume} {33}},\ \bibinfo {pages} {4317} (\bibinfo
  {year} {1986})}\BibitemShut {NoStop}%
\bibitem [{\citenamefont {Zandt}\ \emph {et~al.}(2007)\citenamefont {Zandt},
  \citenamefont {Dwelk}, \citenamefont {Janowitz},\ and\ \citenamefont
  {Manzke}}]{Zandt}%
  \BibitemOpen
  \bibfield  {author} {\bibinfo {author} {\bibfnamefont {T.}~\bibnamefont
  {Zandt}}, \bibinfo {author} {\bibfnamefont {H.}~\bibnamefont {Dwelk}},
  \bibinfo {author} {\bibfnamefont {C.}~\bibnamefont {Janowitz}}, \ and\
  \bibinfo {author} {\bibfnamefont {R.}~\bibnamefont {Manzke}},\ }\href@noop {}
  {\bibfield  {journal} {\bibinfo  {journal} {Journal of Alloys and Compounds}\
  ,\ \bibinfo {pages} {216}} (\bibinfo {year} {2007})}\BibitemShut {NoStop}%
\bibitem [{\citenamefont {Murray}\ and\ \citenamefont {Yoffe}(1972)}]{Murray}%
  \BibitemOpen
  \bibfield  {author} {\bibinfo {author} {\bibfnamefont {R.~B.}\ \bibnamefont
  {Murray}}\ and\ \bibinfo {author} {\bibfnamefont {A.~D.}\ \bibnamefont
  {Yoffe}},\ }\href@noop {} {\bibfield  {journal} {\bibinfo  {journal} {Journal
  of Physics C: Solid State Physics}\ }\textbf {\bibinfo {volume} {5}},\
  \bibinfo {pages} {3038} (\bibinfo {year} {1972})}\BibitemShut {NoStop}%
\bibitem [{\citenamefont {Zhang}\ \emph {et~al.}(2012)\citenamefont {Zhang},
  \citenamefont {Wan}, \citenamefont {Wang},\ and\ \citenamefont
  {Koumoto}}]{Zhang2012}%
  \BibitemOpen
  \bibfield  {author} {\bibinfo {author} {\bibfnamefont {R.-z.}\ \bibnamefont
  {Zhang}}, \bibinfo {author} {\bibfnamefont {C.-l.}\ \bibnamefont {Wan}},
  \bibinfo {author} {\bibfnamefont {Y.-f.}\ \bibnamefont {Wang}}, \ and\
  \bibinfo {author} {\bibfnamefont {K.}~\bibnamefont {Koumoto}},\ }\href
  {http://xlink.rsc.org/?DOI=c2cp42949g} {\bibfield  {journal} {\bibinfo
  {journal} {Physical Chemistry Chemical Physics}\ }\textbf {\bibinfo {volume}
  {14}},\ \bibinfo {pages} {15641} (\bibinfo {year} {2012})}\BibitemShut
  {NoStop}%
\bibitem [{\citenamefont {Kim}, \citenamefont {Park},\ and\ \citenamefont
  {Marzari}(2016)}]{Kim2016}%
  \BibitemOpen
  \bibfield  {author} {\bibinfo {author} {\bibfnamefont {T.~Y.}\ \bibnamefont
  {Kim}}, \bibinfo {author} {\bibfnamefont {C.-h.}\ \bibnamefont {Park}}, \
  and\ \bibinfo {author} {\bibfnamefont {N.}~\bibnamefont {Marzari}},\
  }\href@noop {} {\bibfield  {journal} {\bibinfo  {journal} {Nano Letters}\
  }\textbf {\bibinfo {volume} {16}},\ \bibinfo {pages} {2439} (\bibinfo {year}
  {2016})}\BibitemShut {NoStop}%
\bibitem [{\citenamefont {Park}\ \emph {et~al.}(2014)\citenamefont {Park},
  \citenamefont {Bonini}, \citenamefont {Sohier}, \citenamefont {Samsonidze},
  \citenamefont {Kozinsky}, \citenamefont {Calandra}, \citenamefont {Mauri},\
  and\ \citenamefont {Marzari}}]{Park2014}%
  \BibitemOpen
  \bibfield  {author} {\bibinfo {author} {\bibfnamefont {C.-h.}\ \bibnamefont
  {Park}}, \bibinfo {author} {\bibfnamefont {N.}~\bibnamefont {Bonini}},
  \bibinfo {author} {\bibfnamefont {T.}~\bibnamefont {Sohier}}, \bibinfo
  {author} {\bibfnamefont {G.}~\bibnamefont {Samsonidze}}, \bibinfo {author}
  {\bibfnamefont {B.}~\bibnamefont {Kozinsky}}, \bibinfo {author}
  {\bibfnamefont {M.}~\bibnamefont {Calandra}}, \bibinfo {author}
  {\bibfnamefont {F.}~\bibnamefont {Mauri}}, \ and\ \bibinfo {author}
  {\bibfnamefont {N.}~\bibnamefont {Marzari}},\ }\href@noop {} {\bibfield
  {journal} {\bibinfo  {journal} {Nano Letters}\ }\textbf {\bibinfo {volume}
  {14}},\ \bibinfo {pages} {1113} (\bibinfo {year} {2014})}\BibitemShut
  {NoStop}%
\bibitem [{\citenamefont {Efetov}\ and\ \citenamefont {Kim}(2010)}]{kim}%
  \BibitemOpen
  \bibfield  {author} {\bibinfo {author} {\bibfnamefont {D.~K.}\ \bibnamefont
  {Efetov}}\ and\ \bibinfo {author} {\bibfnamefont {P.}~\bibnamefont {Kim}},\
  }\href@noop {} {\bibfield  {journal} {\bibinfo  {journal} {Physical Review
  Letters}\ }\textbf {\bibinfo {volume} {105}},\ \bibinfo {pages} {256805}
  (\bibinfo {year} {2010})}\BibitemShut {NoStop}%
\bibitem [{\citenamefont {Fratini}\ and\ \citenamefont
  {Guinea}(2008)}]{Fratini2008}%
  \BibitemOpen
  \bibfield  {author} {\bibinfo {author} {\bibfnamefont {S.}~\bibnamefont
  {Fratini}}\ and\ \bibinfo {author} {\bibfnamefont {F.}~\bibnamefont
  {Guinea}},\ }\href@noop {} {\bibfield  {journal} {\bibinfo  {journal}
  {Physical Review B}\ }\textbf {\bibinfo {volume} {77}},\ \bibinfo {pages}
  {195415} (\bibinfo {year} {2008})}\BibitemShut {NoStop}%
\bibitem [{\citenamefont {Morozov}\ \emph {et~al.}(2008)\citenamefont
  {Morozov}, \citenamefont {Novoselov}, \citenamefont {Katsnelson},
  \citenamefont {Schedin}, \citenamefont {Elias}, \citenamefont {Jaszczak},\
  and\ \citenamefont {Geim}}]{Morozov2008}%
  \BibitemOpen
  \bibfield  {author} {\bibinfo {author} {\bibfnamefont {S.~V.}\ \bibnamefont
  {Morozov}}, \bibinfo {author} {\bibfnamefont {K.~S.}\ \bibnamefont
  {Novoselov}}, \bibinfo {author} {\bibfnamefont {M.~I.}\ \bibnamefont
  {Katsnelson}}, \bibinfo {author} {\bibfnamefont {F.}~\bibnamefont {Schedin}},
  \bibinfo {author} {\bibfnamefont {D.~C.}\ \bibnamefont {Elias}}, \bibinfo
  {author} {\bibfnamefont {J.~A.}\ \bibnamefont {Jaszczak}}, \ and\ \bibinfo
  {author} {\bibfnamefont {A.~K.}\ \bibnamefont {Geim}},\ }\href@noop {}
  {\bibfield  {journal} {\bibinfo  {journal} {Physical Review Letters}\
  }\textbf {\bibinfo {volume} {100}},\ \bibinfo {pages} {016602} (\bibinfo
  {year} {2008})}\BibitemShut {NoStop}%
\bibitem [{\citenamefont {Singh}\ \emph {et~al.}(2015)\citenamefont {Singh},
  \citenamefont {Neek-Amal}, \citenamefont {Costamagna},\ and\ \citenamefont
  {Peeters}}]{singh}%
  \BibitemOpen
  \bibfield  {author} {\bibinfo {author} {\bibfnamefont {S.~K.}\ \bibnamefont
  {Singh}}, \bibinfo {author} {\bibfnamefont {M.}~\bibnamefont {Neek-Amal}},
  \bibinfo {author} {\bibfnamefont {S.}~\bibnamefont {Costamagna}}, \ and\
  \bibinfo {author} {\bibfnamefont {F.~M.}\ \bibnamefont {Peeters}},\
  }\href@noop {} {\bibfield  {journal} {\bibinfo  {journal} {Physical Review
  B}\ }\textbf {\bibinfo {volume} {91}},\ \bibinfo {pages} {014101} (\bibinfo
  {year} {2015})}\BibitemShut {NoStop}%
\bibitem [{\citenamefont {Sulania}\ \emph {et~al.}(2016)\citenamefont
  {Sulania}, \citenamefont {Agarwal}, \citenamefont {Kumar}, \citenamefont
  {Kumar},\ and\ \citenamefont {Kumar}}]{Sulania}%
  \BibitemOpen
  \bibfield  {author} {\bibinfo {author} {\bibfnamefont {I.}~\bibnamefont
  {Sulania}}, \bibinfo {author} {\bibfnamefont {D.~C.}\ \bibnamefont
  {Agarwal}}, \bibinfo {author} {\bibfnamefont {M.}~\bibnamefont {Kumar}},
  \bibinfo {author} {\bibfnamefont {S.}~\bibnamefont {Kumar}}, \ and\ \bibinfo
  {author} {\bibfnamefont {P.}~\bibnamefont {Kumar}},\ }\href {\doibase
  10.1039/C6CP03409H} {\bibfield  {journal} {\bibinfo  {journal} {Physical
  Chemistry Chemical Physics}\ }\textbf {\bibinfo {volume} {18}},\ \bibinfo
  {pages} {20363} (\bibinfo {year} {2016})}\BibitemShut {NoStop}%
\bibitem [{\citenamefont {Sandoval}, \citenamefont {Chen},\ and\ \citenamefont
  {Irwin}(1992)}]{Sandova}%
  \BibitemOpen
  \bibfield  {author} {\bibinfo {author} {\bibfnamefont {S.~J.}\ \bibnamefont
  {Sandoval}}, \bibinfo {author} {\bibfnamefont {X.~K.}\ \bibnamefont {Chen}},
  \ and\ \bibinfo {author} {\bibfnamefont {J.~C.}\ \bibnamefont {Irwin}},\
  }\href@noop {} {\bibfield  {journal} {\bibinfo  {journal} {Physical Review
  B}\ }\textbf {\bibinfo {volume} {45}},\ \bibinfo {pages} {14347} (\bibinfo
  {year} {1992})}\BibitemShut {NoStop}%
\bibitem [{\citenamefont {Zhang}\ \emph {et~al.}(2010)\citenamefont {Zhang},
  \citenamefont {Tan}, \citenamefont {Jiang-BinWu}, \citenamefont {Shi},\ and\
  \citenamefont {Tan}}]{nanoscale}%
  \BibitemOpen
  \bibfield  {author} {\bibinfo {author} {\bibfnamefont {X.}~\bibnamefont
  {Zhang}}, \bibinfo {author} {\bibfnamefont {Q.-H.}\ \bibnamefont {Tan}},
  \bibinfo {author} {\bibnamefont {Jiang-BinWu}}, \bibinfo {author}
  {\bibfnamefont {W.}~\bibnamefont {Shi}}, \ and\ \bibinfo {author}
  {\bibfnamefont {P.-H.}\ \bibnamefont {Tan}},\ }\href@noop {} {\bibfield
  {journal} {\bibinfo  {journal} {Nanoscale}\ ,\ \bibinfo {pages} {1}}
  (\bibinfo {year} {2010})}\BibitemShut {NoStop}%
\bibitem [{\citenamefont {Ghorbani-Asl}\ \emph {et~al.}(2013)\citenamefont
  {Ghorbani-Asl}, \citenamefont {Zibouche}, \citenamefont {Wahiduzzaman},
  \citenamefont {Oliveira}, \citenamefont {Kuc},\ and\ \citenamefont
  {Heine}}]{srep}%
  \BibitemOpen
  \bibfield  {author} {\bibinfo {author} {\bibfnamefont {M.}~\bibnamefont
  {Ghorbani-Asl}}, \bibinfo {author} {\bibfnamefont {N.}~\bibnamefont
  {Zibouche}}, \bibinfo {author} {\bibfnamefont {M.}~\bibnamefont
  {Wahiduzzaman}}, \bibinfo {author} {\bibfnamefont {A.~F.}\ \bibnamefont
  {Oliveira}}, \bibinfo {author} {\bibfnamefont {A.}~\bibnamefont {Kuc}}, \
  and\ \bibinfo {author} {\bibfnamefont {T.}~\bibnamefont {Heine}},\
  }\href@noop {} {\bibfield  {journal} {\bibinfo  {journal} {Scientific
  Reports}\ }\textbf {\bibinfo {volume} {3}},\ \bibinfo {pages} {2961}
  (\bibinfo {year} {2013})}\BibitemShut {NoStop}%
\end{thebibliography}
\end{document}